\documentclass[fleqn,usenatbib]{mnras}

\usepackage{newtxtext,newtxmath}

\usepackage[T1]{fontenc}

\DeclareRobustCommand{\VAN}[3]{#2}
\let\VANthebibliography\thebibliography
\def\thebibliography{\DeclareRobustCommand{\VAN}[3]{##3}\VANthebibliography}

\usepackage{graphicx}	
\usepackage{amsmath}	
\usepackage{mathtools}

\DeclarePairedDelimiter\floor{\lfloor}{\rfloor}                      

\usepackage{color,soul}

\title[Tree-based hybrid RHD scheme]{Hybrid radiation hydrodynamics scheme with adaptive gravity-tree-based pseudo-particles}

\author[C. S. C. Lau et al.]{
Cheryl S. C. Lau,$^{1}$\thanks{E-mail: cscl1@st-andrews.ac.uk}
Maya A. Petkova,$^{2}$
and Ian A. Bonnell$^{1}$
\\
$^{1}$School of Physics and Astronomy, University of St Andrews, North Haugh, St Andrews KY16 9SS, UK\\
$^{2}$Department of Space, Earth and Environment, Chalmers University of Technology, SE-412 96 Gothenburg, Sweden
}

\date{Accepted XXX. Received YYY; in original form ZZZ}

\pubyear{2024}

\begin{document}
\label{firstpage}
\pagerange{\pageref{firstpage}--\pageref{lastpage}}
\maketitle

\begin{abstract}
H {\scshape ii} regions powered by ionizing radiation from massive stars drive the dynamical evolution of the interstellar medium. Fast radiative transfer methods for incorporating photoionization effects are thus essential in astrophysical simulations. Previous work by Petkova et al. established a hybrid radiation hydrodynamics (RHD) scheme that couples Smoothed Particle Hydrodynamics (SPH) to grid-based Monte Carlo Radiative Transfer (MCRT) code. This particle-mesh scheme employs the Exact mapping method for transferring fluid properties between SPH particles and Voronoi grids on which the MCRT simulation is carried out. The mapping, however, can become computationally infeasible with large numbers of particles or grid cells. We present a novel optimization method that adaptively converts gravity tree nodes into pseudo-SPH particles. These pseudo-particles act in place of the SPH particles when being passed to the MCRT code, allowing fluid resolutions to be temporarily reduced in regions which are less dynamically affected by radiation. A smoothing length solver and a neighbour-finding scheme dedicated to tree nodes have been developed. We also describe the new heating and cooling routines implemented for improved thermodynamic treatment. We show that this tree-based RHD scheme produces results in strong agreement with benchmarks, and achieves a speed-up that scales with the reduction in the number of particle-cell pairs being mapped.  
\end{abstract}




\section{Introduction} \label{sec:introduction}

Stellar photoionization feedback imposes substantial impact on the properties of molecular clouds. On one hand, the expansion of H {\scshape ii} regions due to thermal overpressure could sweep surrounding gas into shells or pillars, promoting star formation within these self-gravitating structures \citep[e.g.][]{elmegreenlada77,whitworth94a,dale07a,dale09,dalebonnell12,walch13}. Meanwhile, the heating and dispersal of gas from embedded regions by the ionization-driven outflows could halt further fragmentation \citep[e.g.][]{bate09,dale12b,geen17,ali18}. In any case, ionizing radiation serves as a vital ingredient in star formation models. Quantifying the impact of radiative feedback from massive stars is crucial to building up a bigger picture of the galactic ecosystems. 

Over the past decades, our understanding of H {\scshape ii} nebulae physics \citep[e.g.][]{kahn54,osterbrock74} gave rise to analytical solutions that describe the evolution of ionized regions \citep[e.g.][]{spitzer78}. Nonetheless, density structures around young massive stars are typically highly anisotropic due to turbulence, accretion flows, or the gravitational amplifications of fluctuations in the density field \citep[e.g.][]{dalebonnell11}. Given the complexity of local environments in star-forming regions, numerical calculations of radiative transfer (RT) must be relied upon to incorporate their effects into hydrodynamical simulations. 

Modelling the coupling between matter and radiation is the essence of RT algorithms. Early techniques involve moment-based methods, whereby the radiation field is integrated into the hydrodynamic equations and solved using the flux-limited diffusion (FLD) method \citep[e.g.][]{whitehouse05,krumholz07a}. An issue, however, is that the diffusion approximation becomes inappropriate in regions with sharp opacity gradients, such as the feedback-driven cavity shells \citep[][]{kuiper12}, which are ubiquitous around H {\scshape ii} bubbles. The other commonly used technique is ray-tracing, where RT equations are solved directly on lines drawn between the sources and the surrounding fluid elements \citep[e.g.][]{kesseldeynetburkert00,dale07c}. These rays may split recursively for spatial refinement \citep[e.g.][]{bisbas09,baczynski15}. 

Whilst these techniques had been vastly proven to be accurate and effective, one shortcoming is that they are unable to model the photon scattering, absorption and re-emission. Neglecting these processes limits our ability to understand the effect of dust on H {\scshape ii} regions \citep[e.g.][]{haworth15,ali20}. One could argue that such stochasticity at atomic level may be neglected from the large-scale perspective, but plenty astrophysical phenomena, such as the production of ISM diffuse field \citep[e.g.][]{haworthharries12} or the reprocessing of UV photons \citep[e.g.][]{witt92}, are crucial pieces of physics which ray-tracing methods cannot fully take into account. To the contrary, the Monte Carlo radiative transfer (MCRT) technique overcomes these limitations by explicitly simulating the random events.

MCRT is a grid-based method that works by discretizing the radiative source into photon packets, then propagating them through a given density field to perform individual random walks. Meanwhile, each grid cell accumulates the packets' path lengths and estimates the steady-state ionic fraction. The packet-emission is repeated until the temperature and ionization structures converge. For further details, the interested reader is referred to e.g. \citet{och98,ercolano03,wood04}\footnote{We note that gridless MCRT methods have also been developed \citep[e.g.][]{pawlikschaye08,altay08} (see Section~\ref{sec:gridless_mcrt}), but these methods rely on alternative approaches to calculate photoionization rate that is beyond our scope.}. Despite being computationally expensive, the MCRT method holds great advantage in handling radiation self-consistently even in highly inhomogeneous environments. 

In view of this, \citet{petkova21} presented a radiation hydrodynamics (RHD) scheme that couples an MCRT ionization code to a smoothed particle hydrodynamics \citep[SPH; ][]{gingoldmonaghan77,lucy77} code. MCRT methods are traditionally used for post-processing hydro simulation snapshots to produce synthetic emission maps \citep[e.g.][]{lomaxwhitworth16}, and so far adopting MCRT as a `live feedback' computation is still relatively uncommon. A current notable example is the live coupling of SPH code {\scshape phantom} \citep[][]{phantom18} to the MCRT code {\scshape mcfost} \citep{pinte06,pinte09} for simulating dust-processed radiation and computing gas temperatures in protoplanetary discs \citep[e.g.][]{pinte19,nealon20,borchert22}. In a similar vein, our RHD scheme couples {\scshape phantom} to the MCRT code {\scshape cmacionize} \citep[][]{vandenbrouckewood18,vandenbrouckecamps20} for modelling photoionization. 

The RHD scheme works as follows. At each hydro step in {\scshape phantom}, the SPH fluid density field is passed to {\scshape cmacionize} to execute the MCRT simulation on a Voronoi grid \citep[][]{voronoi1908} with initial generation sites coinciding with the particles' locations. The MCRT simulation solves for the steady-state ionic fraction of each cell, and the results are subsequently returned to {\scshape phantom}. With this, we heat up the ionized SPH particles, thus exert a thermodynamical impact on to the fluid. 

Due to the meshless nature of SPH, coupling it to a grid-based MCRT code requires transferring fluid properties between the two models. This RHD scheme uses the \textit{Exact} density mapping method, developed by \citet{petkova18}, that allows particle-interpolated densities to be accurately reconstructed on a Voronoi grid with which the MCRT simulation is run. The Exact mapping is formulated based on an analytical solution to the 3D volume-integral of SPH interpolation for a Voronoi cell of any geometry \citep[][]{petkova18}. This method allows the Voronoi grid to be modified independently of the particles whilst ensuring mass conservation. 

However, in spite of its excellent accuracy, a major drawback of this RHD scheme is the computing time. Both the MCRT simulation and the Exact density mapping method are highly expensive procedures. As runtime rises log-log linearly with the number of particle-cell pairs (cf. Appendix A in \citet{petkova21}), the simulations become infeasible with over $10^5$ SPH particles, rendering it undesirable for practical applications. This urges the need to improve the algorithm’s efficiency in the large particle number regime, allowing this RHD scheme to operate for large scale numerical experiments. 

This paper is organized as follows. In Section \ref{sec:methods}, we describe our novel tree-based optimization method developed to accelerate this RHD scheme. We emphasize that this algorithm is applicable to any astrophysical SPH codes where gravity trees are in place. We also introduce in Section \ref{sec:heating_cooling_method} the heating and cooling computation methods for improving the thermal physics treatments. In Section \ref{sec:results}, we present the test results to illustrate the accuracy and benefits of the new implementations. Finally, we discuss the applications in Section \ref{sec:applications} and summarize in Section \ref{sec:conclusion}.

\section{Numerical methods} \label{sec:methods}

\subsection{Tree-based radiation algorithms} \label{sec:tree_rt}

The radiation transport problem for \textit{N} fluid resolution elements has an intrinsic complexity of $\mathcal{O} ( N^{7/3} )$ \citep[cf. e.g.][]{grond19}. Developing low-scaling algorithms to model radiation fields has been one of the priorities in numerical astrophysics. Recently there has been a growing tendency to utilize the tree-based gravity solvers (known as \textit{gravity trees}) into RT methods. Gravity trees refer to the tree data structures that hierarchically group the fluid resolution elements, with each group known as a \textit{node}. The root node represents the whole simulation domain. These methods are developed to solve self-gravity in a computationally feasible manner. Consider a target element upon which we evaluate the total gravitational force. By traversing (i.e. \textit{walking}) the tree with respect to this target, the distant elements may be collectively treated as larger fluid parcels, such that their force contributions can be approximated using multipole expansions rather than calculated via a direct summation, reducing the complexity from $\mathcal{O} ( N^2 )$ to $\mathcal{O} ( N \log N )$ \citep[e.g.][]{barneshut86}. The way how the tree is walked determines the sizes of these parcels and thus the computation accuracy of gravity; tree traversals are controlled by the node \textit{opening criteria} (see Section~\ref{sec:treewalk}). 

Gravity trees have been adopted in astrophysics since the development of hydrodynamical simulations. Some notable examples include the ‘Press’ tree \citep[][]{press86,benz88} used in {\scshape sphng} \citep[][]{benz90}, the {\scshape treesph} \citep[][]{hernquistkatz89} used in {\scshape gadget-2} \citep[][]{springel05}, the Barnes-Hut tree \citep[][]{barneshut86} used in {\scshape arepo} \citep[][]{arepo20}, and the \textit{k}-d tree \citep[][]{bentley79} used in {\scshape gasoline} \citep[][]{wadsley04}, covering both particle-based SPH/\textit{N}-body codes and grid-based Adaptive Mesh Refinement codes. The availability of  gravity trees in astrophysical simulation codes makes them highly suitable for RT applications, since they already keep track of the distribution of matter in the simulation domain. 

Trees enable a fast retrieval of information on the intervening materials along the photon paths without extra computation overhead. For example, the {\scshape treecol} algorithm developed by \citet{clark12} uses the gravity tree to create angular distribution maps of column densities as seen from the fluid elements for measuring the self-shielding of gas \citep[e.g.][]{smith14}. Later, this evolved into the {\scshape treeray} algorithm by \citet{wunsch18,wunsch21} and \citet{haid18}. {\scshape treeray} combines the gravity tree with the reverse ray-tracing method, whereby the radiative sources are all embedded into the tree during tree-build, allowing the distant sources to be `merged' with respect to the target fluid elements. In both algorithms, the physical properties for RT are computed \textit{whilst} the self-gravity is solved. However, like the gravitational force method, {\scshape treeray} walks the tree only with pre-defined criteria, and is unable to adapt to the actual amount of flux received by the fluid element. This issue necessitates the implementation of adaptive tree-walk systems in tree-based RT, as demonstrated by \citet{grond19} in their {\scshape trevr} code, which ensure that the regions with steep density gradients (such as in a clumpy medium) are adequately resolved. 

Furthermore, the MCRT technique itself has a long history of adopting tree-based adaptive refinement approaches. Tree data structures are used in grid-partitioning to resolve regions where dust fraction is high, or where radiation field varies rapidly, hence achieving the most efficient use of computing power \citep[][]{wolf99,kurosawahillier01,saftly14}. In this work, employing adaptive grids for the MCRT simulation is indeed a plausible way to speed up the calculations. However, because of the density mapping, the computation would still be overwhelmed by the large particle number, let alone that tailored grids are difficult to be set up \textit{before} the MCRT without a good prior estimate of the radiation field. 

For this, we ought to treat the coupling itself. In the same vein as the tree-based methods described above, we apply the gravity tree to `merge' fluid elements in regions which are less dynamically affected by the ionizing radiation. Here, the most intuitive way to optimize an SPH-MCRT coupled scheme is to reduce the number of particles `seen' by the MCRT code. Since the RT is not computed within SPH in this scheme, we can simply convert the gravity tree nodes into pseudo-SPH particles\footnote{The name `pseudo-particle' has also been used to describe gravity tree nodes in \citealp[e.g.][]{dale14,schafer16}.}, whose positions, masses and smoothing lengths are then used to compute optical depths in the MCRT simulation (see Section~\ref{sec:pseudo_particles} for details). Similar to \citet{grond19}, we also implement an adaptive tree-walk algorithm to ensure that regions of high ionic fractions are sufficiently well resolved by the MCRT grid. Thus, this pseudo-particle method is, in essence, a tree-based adaptive refinement system which is dedicated to particle-mesh coupled codes. 

\subsection{The \textit{k}-d gravity tree} \label{sec:kd_tree}

We first provide an overview of the gravity tree used in our RHD scheme, though we stress that the methods are generalizable to other tree-based gravity solvers. The tree currently implemented in {\scshape phantom} is a binary \textit{k}-d tree, and is developed based on the recursive coordinate bisection (RCB) tree algorithm proposed by \citet{gaftonrosswog11}, which was specifically designed for astrophysical SPH codes. This \textit{k}-d tree is highly efficient for both gravitational computation and neighbour-search, with the latter being crucial for computing SPH kernels. Details on the tree-build and tree-walk algorithms are documented in \citet{phantom18}, however the reader is reminded that modifications have been made since the publication of the paper. We describe below the main features of this tree. 

\subsubsection{Tree-build} \label{sec:treebuild}

\begin{figure}
    \centering
    \includegraphics[width=2.7in]{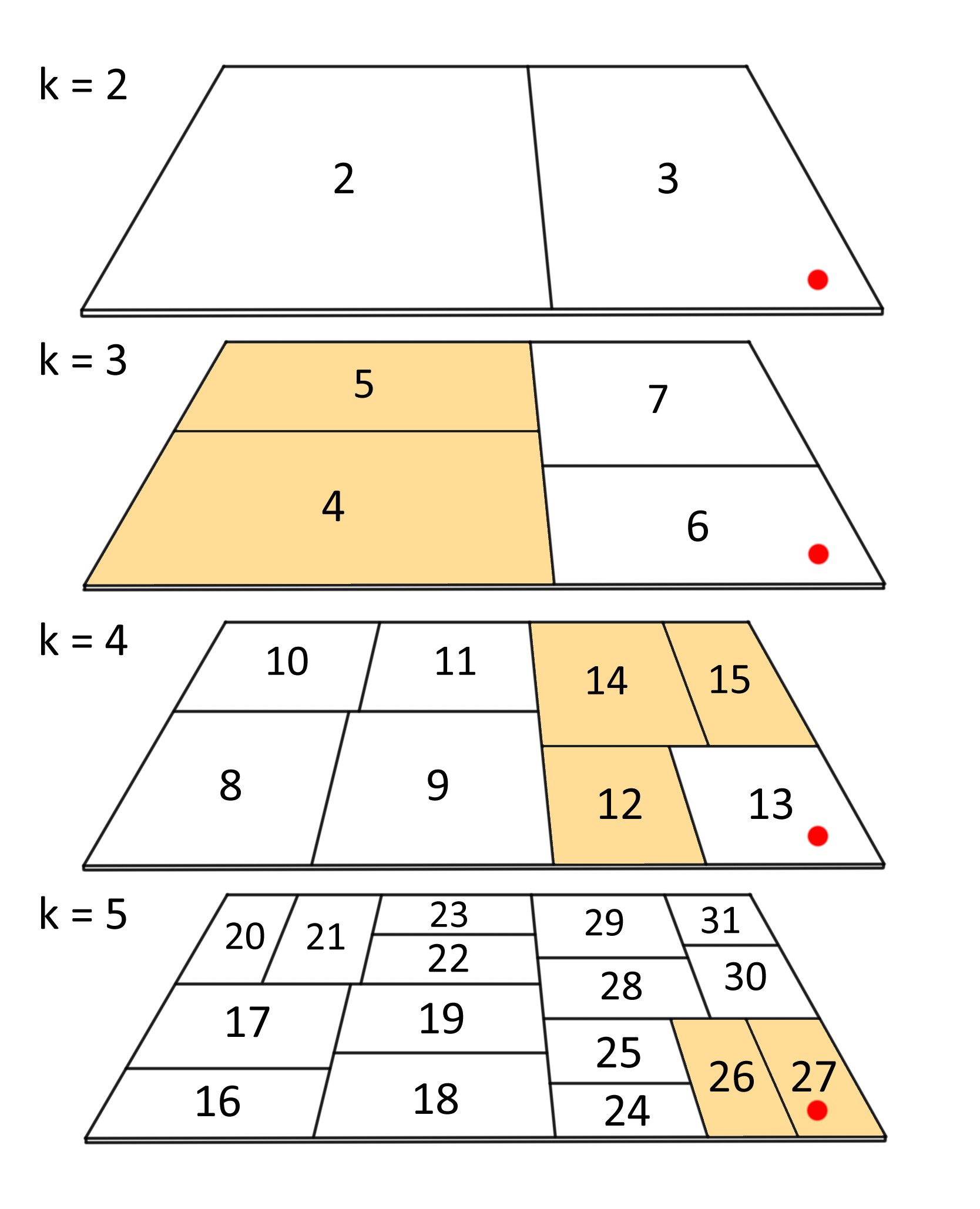}
    \caption{Illustration of the recursive splitting of simulation domain to build the \textit{k}-d gravity tree. The 3-D space is represented as a 2-D plane, and each plane in the stack represents a tree level. The sub-domains (nodes) are labelled according to the convention of \citet{gaftonrosswog11} (see also Fig.~\ref{fig:tree_structure}). For illustration purpose, tree-build is terminated at the $5^\mathrm{th}$ level. Consider a target particle at the position indicated by the red dot. Only the coloured nodes are considered in its gravitational force evaluation (see Section~\ref{sec:treewalk}).}
    \label{fig:tree_layers}
\end{figure}

\begin{figure}
    \centering
    \includegraphics[width=3.3in]{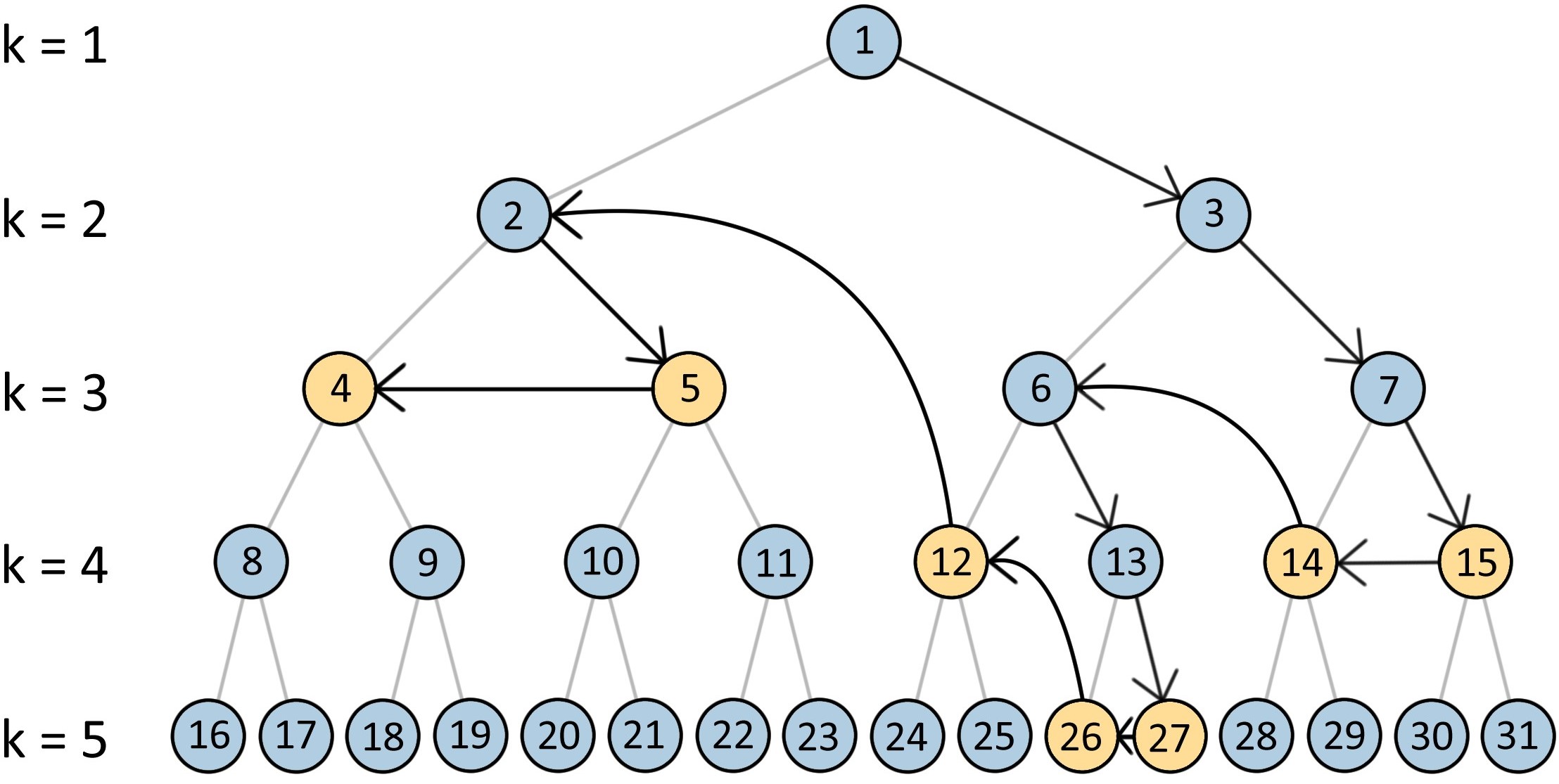}
    \caption{Structure of the \textit{k}-d tree that corresponds to the example shown in Fig.~\ref{fig:tree_layers}. The labelling convention enables node relations to be easily recovered. Arrows indicate an example tree-walk sequence with opening criteria defined with respect to the target particle marked in Fig.~\ref{fig:tree_layers} (see Section~\ref{sec:treewalk}). Nodes in orange are those which would have been \textit{accepted} during this traversal. Their physical domains are also coloured in orange in Fig. \ref{fig:tree_layers}. }
    \label{fig:tree_structure}
\end{figure}

The \textit{k}-d tree adopts a `top-down' approach in the building process. Starting from the root node, the nodes’ spatial domains are recursively split into two. The split is through the centre of mass of the constituting particles to maintain load balancing, and along the longest axis of the cell to avoid elongation. Each split adds a level to the tree and the tessellated cells become the daughter nodes. This procedure is repeated until the leaf nodes (often referred to as `lowest-level cells') contain less than a certain number of particles. In {\scshape phantom}, by default, the splitting ends when the leaf node contains less than 10 particles. We illustrate the tree structure in Fig.~\ref{fig:tree_layers} and Fig.~\ref{fig:tree_structure}. Position of a node is defined to be its centre of mass, and size $s_\mathrm{node}$ is defined to be the distance from its centre of mass to its furthest constituent particle. Each node also stores the computed quadrupole moments for evaluating the long-range gravitational accelerations \citep[e.g.][]{benz90}, though this quantity is not required in our RHD scheme. 

Following \citet{gaftonrosswog11}, the nodes are labelled such that their indices carry information of local proximity. The root is given an index of 1, and the subsequent nodes are labelled in ascending numerical order, from left to right, progressively down the tree. An example is demonstrated in Fig.~\ref{fig:tree_structure}. This labelling convention gives rise to a set of convenient arithmetic rules that allow node relations to be recovered without extra memory consumption. For instance, a node with index $n_a$ is located on level $k_a \equiv \floor*{\log_{2}{n_a}}$; its parent $n_b$ on some higher level $k_b$ can be determined with the formula $n_b = \floor*{n_a / 2^{k_a - k_b}}$. We utilize these relations in Section~\ref{sec:neighfind} for our neighbour-find algorithm. Note that because the tree rebuilds as particles evolve, the node labels that correspond to each spatial domain in the simulation likely change across timesteps. 

Traditionally, gravity trees in astrophysical SPH codes are built in a `bottom-up' manner.  An example is the `Press' tree in {\scshape sphng}, where particles are being hierarchically grouped together until reaching the root node. In this case, the leaf nodes \textit{are} the individual SPH particles. Top-down approaches, in comparison to bottom-up methods, grant the freedom to create `larger' leaf nodes. It allows the long-range gravitational force to be evaluated once per lowest-level cell rather than once per particle, hence achieving a much higher efficiency at a small cost in accuracy. But whilst it is more advantageous for self-gravity computations, in Section~\ref{sec:adaptive_tree_walk} we describe the additional treatments required in setting up the pseudo-particles due to this aspect of top-down trees. 

\subsubsection{Tree-walk} \label{sec:treewalk}

The tree is subsequently traversed for each lowest-level cell (or any target point) to evaluate the total gravitational force acting on it. Here, we describe the depth-first search algorithm implemented in {\scshape phantom}. The arrows in Fig.~\ref{fig:tree_structure} illustrate an example traversal. Tree-walk begins from the root. We go through a series of opening criteria to decide whether or not this node needs to be resolved into its constituents. If yes, we examine its child node on the right. This downward search is continued until the opening criterion is no longer met, we \textit{accept} the node as it is, and compute its force contribution via a multipole expansion. Thereupon, we turn left and look for the adjacent branch that is yet to be traversed. The tree-walk is terminated when all branches have been visited. In Fig.~\ref{fig:tree_structure} and Fig.~\ref{fig:tree_layers}, nodes coloured in orange are those which would have been extracted in this example traversal to compute gravity. The closer to the target, the lower it is on the tree, and vice versa. 

It is trivial to see that the physical domains of the accepted nodes extracted from tree-walks are effectively a re-tessellation of the simulation space, implying that total mass must be conserved. Like SPH particles, nodes also represent fluid parcels, only of different sizes and masses. Hence, converting these nodes into pseudo-particles does not modify the underlying density distribution, but only their resolution lengths. As such, tree-walks serve as a great tool for adjusting the resolution of the fluid elements as seen by the MCRT code. 

\subsection{Code overview} \label{sec:code_overview}

\begin{figure*}
    \includegraphics[width=7in]{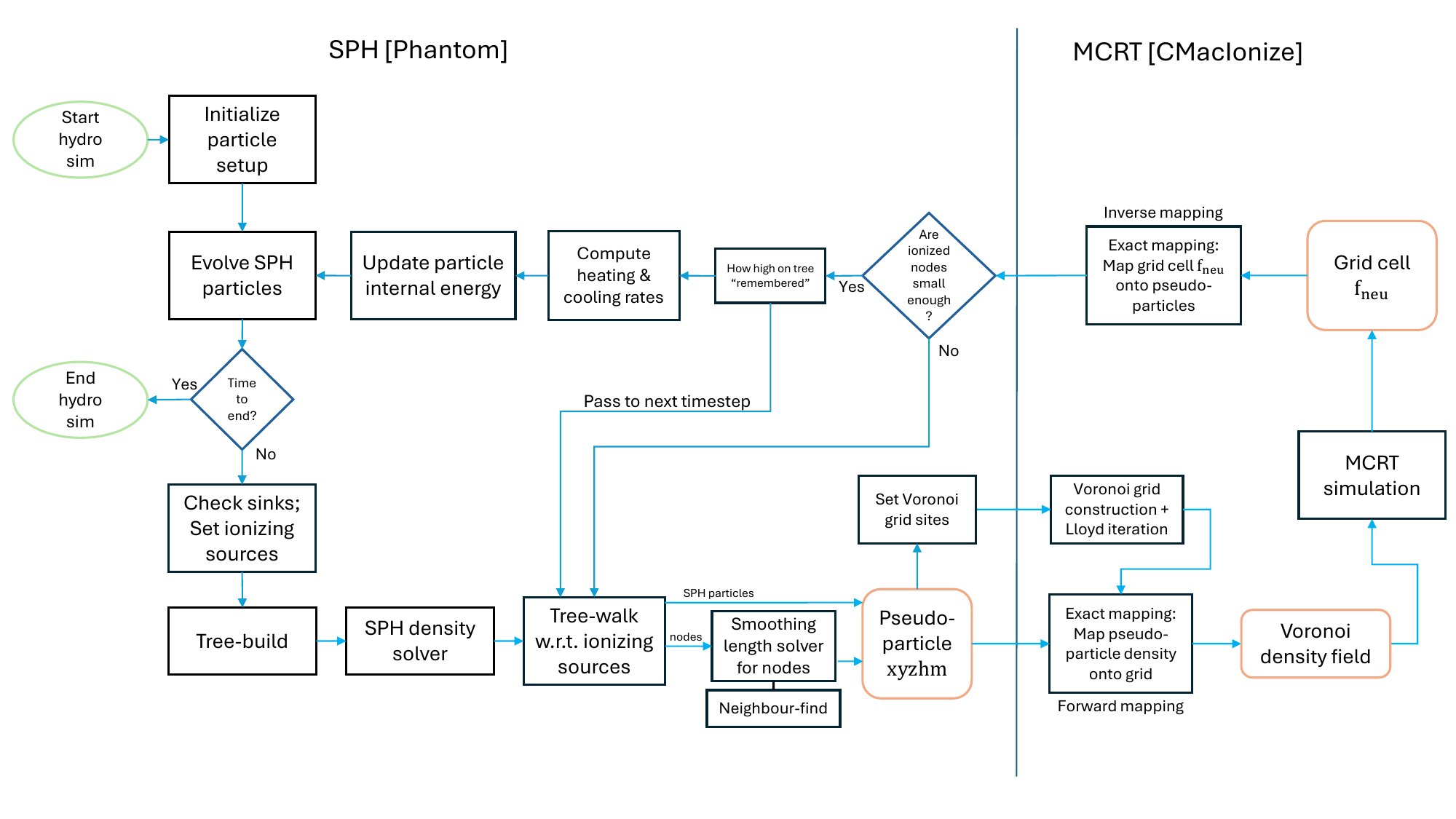}
    \caption{Flowchart showing the main computation procedures involved in this RHD scheme. Procedures on the left-hand side are those executed within the SPH code {\scshape phantom}, and those on the right-hand side are carried out within the MCRT code {\scshape cmacionize}. The vertical line indicates the interface between the two codes, which is established using the {\scshape fortran} library in {\scshape cmacionize}. The abbreviation xyzhm denotes the positions, smoothing lengths and masses of the pseudo-particles, and $f_\mathrm{neu}$ is the neutral fraction (= 1 - ionic fraction $f_\mathrm{ion}$). }
    \label{fig:code_overview}
\end{figure*}

Before discussing the details on setting up pseudo-particles, this section outlines computation procedures involved in this RHD scheme. We highlight in particular the newly added operations compared to the original version described in \citet{petkova21}. The flow of the algorithm is illustrated in Fig.~\ref{fig:code_overview}. Note also that this RHD scheme has now been implemented as one of the physics modules in {\scshape phantom}, meaning the call to {\scshape cmacionize} and the photoionization heating computations are being integrated into {\scshape phantom}'s Leapfrog stepping algorithm \citep[][]{verlet67,tuckerman92}, rather than through the {\scshape live analysis} module which is typically only used for post-processing simulation snapshots. Doing this enables the timesteps to be promptly constrained by the photoionization heating, in consistent with other heating and cooling operations in {\scshape phantom}. It also means that {\scshape cmacionize} is called every (individual) timestep. For the current implementation, we assume a pure hydrogen gas composition for all calculations in the code. Extensions to treat other atomic species may be carried out in future developments. 

At the beginning of each hydro step, we check the self-consistently formed sink particles \citep[][]{bate95} to locate all ionizing sources. The sinks are assumed to represent individual stars in the current implementation, but extending the code to treat sinks as clusters is straightforward. We assume the ionizing photon frequencies from sinks follow a blackbody spectrum, and we estimate the stellar effective temperature $T_{\ast}$ with the sink mass $M_{\ast}$ using the Mass-Luminosity relation for main sequence and the Stefan-Boltzmann relation. This temperature $T_{\ast}$ is also used for computing the photoionization heating rates (see Section~\ref{sec:heating_method}). To calculate the Lyman continuum photon flux $Q_H$, we use the formula in \citet{dale12b},
\begin{equation}
    \log(Q_H) = 48.1 + 0.02 ( M_{\ast} - 20 M_\odot), 
	\label{eq:ionflux_starmass}
\end{equation}
which is an analytical fit function to the tabulated relation between ionizing photon flux and mass of solar-metallicity stars in \citet[see Table~1]{diazmiller98}. Alternatively, monochromatic sources may be manually set by the user. With the given ionizing photon flux, we estimate the stellar temperature by, again, fitting a function to Table 1 of \citet{diazmiller98}, giving 
\begin{equation}
    T_{\ast} = 0.89 \exp (0.21 \log_{10}{(Q_H)} + 0.28) + 7613.22. 
	\label{eq:temp_ionflux}
\end{equation}

We also remind the reader that this RHD scheme uses {\scshape cmacionize} as a pure time-independent photoionization code, meaning that ionization equilibrium is enforced. As such, it is necessary to ensure that the timestep covers the time taken for the H {\scshape ii} regions to reach photoionization steady-state. This time duration may be approximated as the recombination timescale, 
\begin{equation}
    t_D = \frac{m_p}{\alpha_B \rho_0} \approx 19.6\ \mathrm{yr} \left( \frac{\rho_0}{10^{-20}\ \mathrm{g\ cm^{-3}}} \right)^{-1}
    \label{eq:recomb_timescale}
\end{equation}
\citep[cf. e.g.][equation 3]{bisbas15}, where $m_p$ is the proton mass, $\alpha_B$ is the case-B recombination coefficient (see equation~\ref{eq:caseBrecomb_temp}), and $\rho_0$ is the ambient medium density. We estimate $\rho_0$ by averaging the densities of SPH particles within $1\ \mathrm{pc}$ around the source. For $\rho_0 > 10^{-20}\ \mathrm{g\ cm^{-3}}$, $t_D$ is under 20 years.  

Once an ionizing source is switched on, if the timestep is smaller than $t_D$, the simulation will be halted and the user will be asked to adjust the (maximum) timestep or the initial conditions. Beyond that, we presume the H {\scshape ii} region would remain in ionization equilibrium and timestep-checks are hence no longer required. This assumption is mostly appropriate for modelling embedded OB stars, since (a) the environment around massive stars are sufficiently dense that $t_D$ is not longer than the dynamical timescale, and (b) the ionizing flux from massive stars do not fluctuate. 

We execute the RHD scheme after the tree-build and SPH density solver, in which the particle smoothing lengths have updated. First, we walk the tree with respect to the ionizing source(s), and extract a set of nodes and particles from the tree which is then turned into pseudo-particles (Section~\ref{sec:adaptive_tree_walk}). This set of pseudo-particles' distribution acts as the initial estimate of the spatial extents of the ionized regions. Note that, unlike {\scshape treecol} and {\scshape treeray}, the tree-walk here is \textit{separate} from the gravity solver. Then, we compute the nodes' smoothing lengths (Section~\ref{sec:h_solver}), using a neighbour-find algorithm which is specifically designed for tree nodes (Section~\ref{sec:neighfind}). Prior to calling {\scshape cmacionize}, the pseudo-particles’ positions are used to set up the Voronoi grid generation sites. We also merge the most tightly packed sites to avoid exceeding the precision limit in the Voronoi grid construction within {\scshape cmacionize}. Fortunately, the code typically merges only the sites along the densest filaments where high resolutions are unnecessary, for they are least dynamically affected by ionizing radiation \citep[e.g.][]{dalebonnell11}. Afterwards, we pass the pseudo-particles’ positions, masses and smoothing lengths as well as the Voronoi generation sites to {\scshape cmacionize}. 

{\scshape cmacionize} first applies 5 Lloyd iterations \citep[][]{lloyd82} to regularize the Voronoi grid by correcting the elongated cells. This step is necessary in order to adequately resolve the regions of lower density, which are likely filled with ionized gas \citep[][]{petkova21}. We then apply the Exact density mapping to transfer the SPH density field from the pseudo-particles on to the Voronoi grid (forward mapping). The MCRT simulation is run with the ionizing sources specified by {\scshape phantom}. We set the packet-release procedure to iterate for 10 times and each time $10^6$ photon packets are emitted into the field. 10 times is typically sufficient for the ionic fraction in each grid cell to converge \citep[cf.][Fig. 5]{vandenbrouckewood18}. In fact, {\scshape cmacionize} can self-consistently solve for the steady-state ionic fraction and the temperature in each cell, but if a pure hydrogen composition is used, the temperature is fixed to $10^4\ \mathrm{K}$. Hence, in this RHD scheme, we track and evolve the particles' thermal properties only within the SPH code\footnote{It is acknowledged that the current implementation could provoke slight inconsistencies in temperatures between the SPH and the MCRT code, especially when the particles are yet to reach thermal equilibrium. Passing the current particle temperatures from SPH into the MCRT code to replace the assumed temperature of $10^4\ \mathrm{K}$ will be the subject of future work.}. After the MCRT simulation is completed, the inverse mapping \citep[][]{petkova18,petkova21} is applied to transfer the ionic fractions of the grid cells back on to the pseudo-particles, and return the results back to {\scshape phantom}. 

Of course, with the user-defined tree opening criteria alone, the initial set of pseudo-particles is likely unable to fully well-resolve the ionized regions, particularly in inhomogeneous environments. Hence, it is necessary to adjust the tree-walks using the ionic fractions returned from {\scshape cmacionize}. This adjustment is carried out iteratively until all ionized pseudo-particles are sufficiently small in size, as determined by a checking criterion (Section~\ref{sec:adaptive_tree_walk}). Only after that, we compute the photoionization heating and cooling for each particle, and update their internal energies in the SPH simulation (Section~\ref{sec:heating_cooling_method}) to proceed to the next hydro step. Crucially, this iterative adjustment system allows the pseudo-particles to adapt to the evolving H {\scshape ii} regions. Performing such trial and error process \textit{within} a hydro step eliminates the need to predict the growth of the ionization fronts, without delaying the `response' to the subsequent timestep which hinders the modelling accuracy. 

\subsection{Pseudo-particles} \label{sec:pseudo_particles}

The density mapping requires the positions, masses and smoothing lengths of the SPH particles. We describe in the following the methods to obtain these quantities for the pseudo-particles. 

\subsubsection{Adaptive tree-walk} \label{sec:adaptive_tree_walk}

\begin{figure*}
    \includegraphics[width=5.4in]{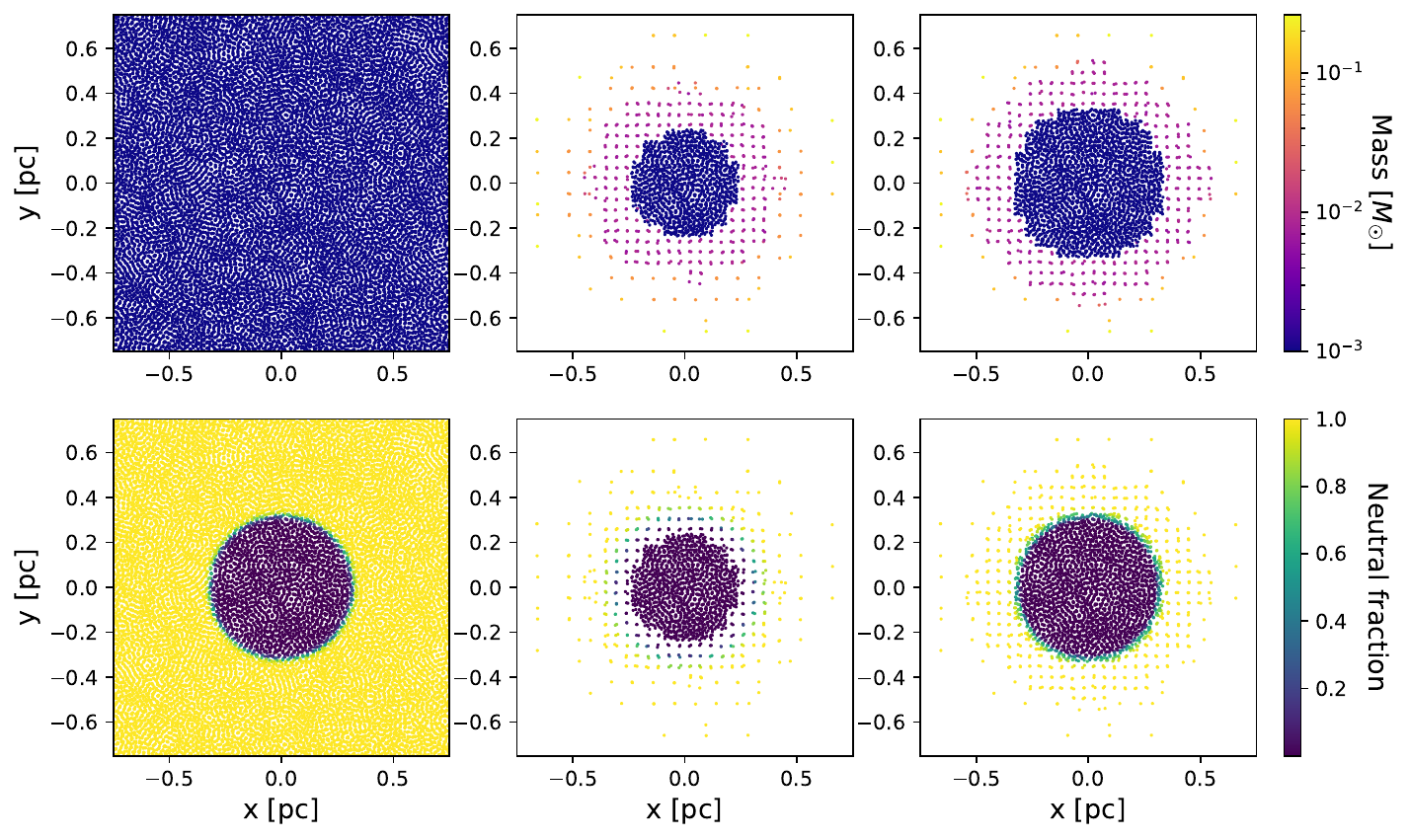}
    \caption{Slices from a 3-D simulation of an ionizing source placed in a uniform medium (see Section~\ref{sec:starbench} for the simulation setup). Left panels illustrate the case with all SPH particles. Middle panels illustrate the pseudo-particles obtained from a tree-walk, with $r_{\mathrm{part}} = 0.2 \ \mathrm{pc}$ and $r_{\mathrm{leaf}} = 0.4 \ \mathrm{pc}$. Right panels show the same as the middle but with the ionization front just resolved after the adaptive tree-walk iterations. Color scales in the upper panels indicate particle mass and bottom panels indicate their neutral fractions. The sharp transition zone between the ionized and the neutral domains defines the ionization front.}
    \label{fig:node_illus}
\end{figure*}

The tree-walk is controlled via multiple opening criteria to set up the pseudo-particles. First, as an initial estimate, we define a threshold radius $r_{\mathrm{part}}$ around the ionizing sources within which we must open the leaves and extract the individual SPH particles. We hereafter consider these SPH particles to be part of the full set of pseudo-particles. We also define another threshold radius $r_{\mathrm{leaf}}$, where $r_{\mathrm{leaf}} > r_{\mathrm{part}}$, such that in the annulus between the two radii, leaf nodes must be extracted. For the outer regions, we follow the criterion used for gravity calculations, that is, to define an opening angle $\theta$, given by 
\begin{equation} 
    \theta^2 <  \left( \frac{s_\mathrm{node}}{r_\mathrm{node}} \right)^2,
    \label{eq:opening_angle}
\end{equation}
where $s_\mathrm{node}$ is the size of the node and $r_\mathrm{node}$ is its distance to the ionizing source. The angle $\theta$ is a parameter set between 0 and 1. If the angle subtended by the node to the source is greater than $\theta$, we open the node. Hence, the smaller the $\theta$, the more pseudo-particles extracted. We typically set $\theta \sim 0.1$. Fig.~\ref{fig:node_illus} shows an example outcome of this tree-walk. The reader may notice that the $r_{\mathrm{leaf}}$ is set to ensure a smooth transition in the pseudo-particle distribution from the particle-level to the higher levels governed by equation~(\ref{eq:opening_angle}). In situations where multiple ionizing sources are present, we traverse the tree only once but allow the opening criteria to take all sources into account. The above procedures provide the first set of pseudo-particles, whose positions and masses are determined from the nodes' properties stored from tree-build, or directly from the SPH particles for those within $r_{\mathrm{part}}$. 

Upon retrieving the ionic fractions returned from {\scshape cmacionize}, the subsequent tree-walks are adjusted, as mentioned in Section~\ref{sec:code_overview}. This is done by checking the size of the ionized pseudo-particles, to see whether or not those with high ionic fractions (low neutral fractions) are sufficiently low on the tree and small in size, hence well-resolved. We parametrize this checking criterion via a function that relates the lower limit in neutral fraction to the relative size of the node, 
\begin{equation} 
    f_{\mathrm{neu,limit}} = \frac{1}{K} \left( -\frac{1}{s_\mathrm{node} / s_\mathrm{root}} + K \right),
    \label{eq:resol_param}
\end{equation}
where $f_{\mathrm{neu,limit}}$ is the neutral fraction threshold, $s_\mathrm{node}$ is the size of the node in concern, $s_\mathrm{root}$ is the size of the root node, and $K$ is a free parameter in the range of $1 < K \lesssim 500$. Nodes with neutral fraction beneath its $f_{\mathrm{neu,limit}}$ are considered under-resolved. We plot equation~(\ref{eq:resol_param}) in Appendix~\ref{appen:tree_resol_K} to show its dependence on the parameter $K$. The higher the $K$ value, the higher the $f_{\mathrm{neu,limit}}$, which tightens the criterion on pseudo-particle size once their neutral fractions drop slightly below 1. This parameter $K$ primarily controls the resolution of the partially ionized zones (see Section~\ref{sec:tree_resolution} for a further discussion). 

If one or more nodes did not pass the checking criterion, we store their labels and open them in the next tree-walk. We iteratively resolve into the ionized regions by `pressing' the pseudo-particles down the tree until all nodes satisfy equation~(\ref{eq:resol_param}). The reader is reminded that this iterative procedure is carried out \textit{within} a hydro step to eliminate time delays in adjusting to the expanded H {\scshape ii} regions. If even the leaves fail, we simply increase $r_{\mathrm{part}}$ (along with $r_{\mathrm{leaf}}$) where no further resolution checks will be performed, since it is impossible to resolve below the particle level. 

\begin{figure}
    \includegraphics[width=3.0in]{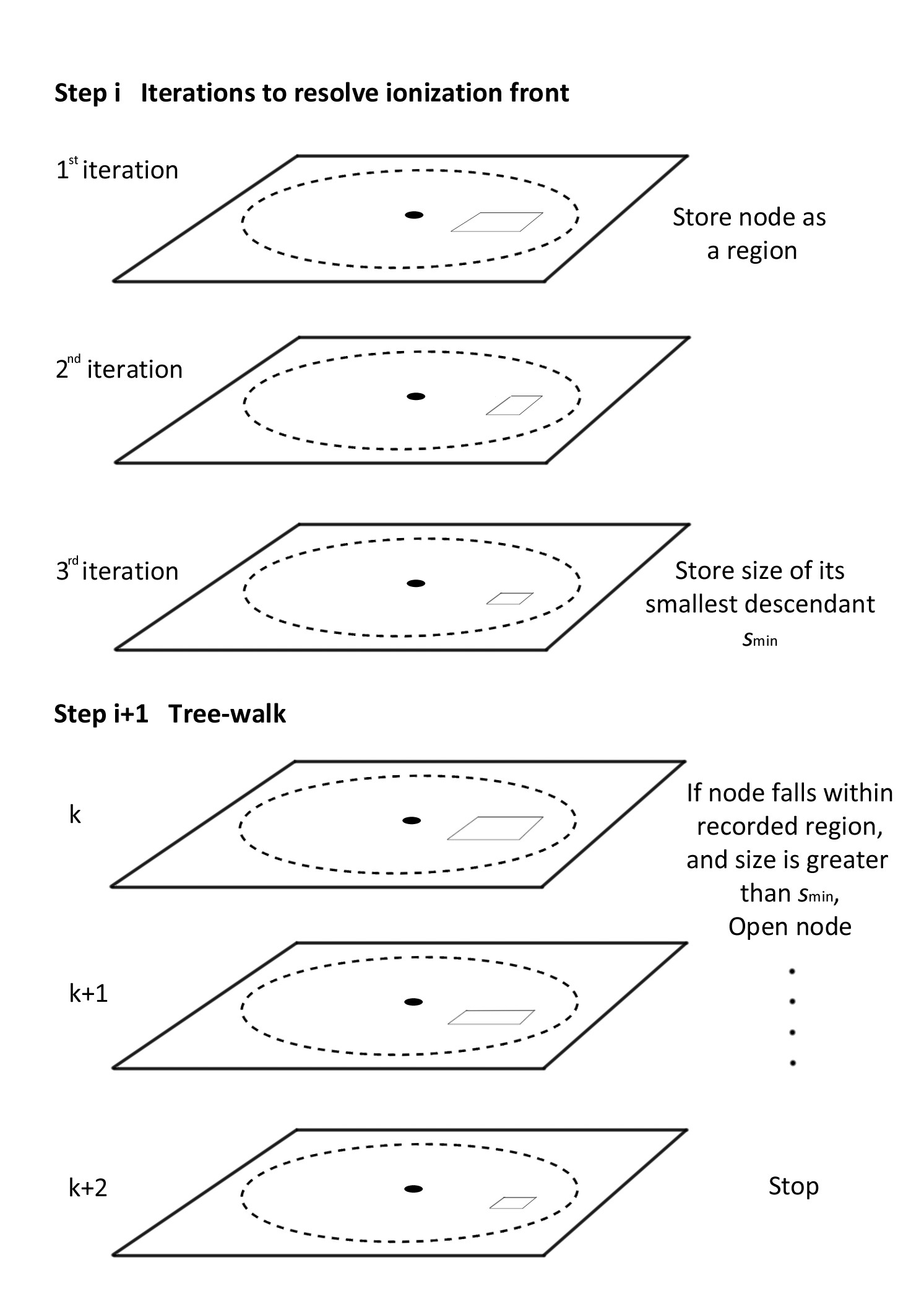}
    \caption{Illustration of method to restore the pseudo-particle distribution after the tree rebuilds. 2-D planes represent the 3-D space. The dotted circle represents the H {\scshape ii} region and the black dot indicates the ionizing source. Grey squares represent arbitrary ionized nodes. Top panel shows the method to record the pseudo-particles' sizes during step $i$, and bottom panel illustrates the method to restore it during tree-walk at step $i+1$. }
    \label{fig:store_tree}
\end{figure}

To avoid repeating these iterations in subsequent steps, an algorithm was designed to let the code `remember' how high the pseudo-particles were on the tree. Keeping in mind that trees would rebuild as particles move, we ought to use the nodes' \textit{physical} properties rather than their labels. The method is depicted in Fig.~\ref{fig:store_tree}. Consider a node that failed the checks during the initial trial, as marked by the grey square on the first plane in the top panel. This node effectively draws a spatial region where higher resolutions are required, and we record it. As the node iteratively resolves into its constituents, we, in the meantime, record the minimum size of its descendants crossed by the tree-walks, as marked by the grey square on the third plane. In the next timestep, if a node (grey square on first plane in bottom panel) falls within the boundaries of this recorded region, we open it and advance down the tree until the node size becomes comparable to that of its previously-recorded smallest descendant. 

This algorithm does not guarantee reproducing the previous set of pseudo-particles, but it is capable of restoring their overall distribution and immediately suppressing the iterations. Hence, this iterative procedure takes place typically only during the first timestep where the pseudo-particles adjust to locate the ionization front. Afterwards, it iterates only when the boundary of the H {\scshape ii} region has moved beyond the resolved area. If an ionized node becomes shielded from the stellar source, we simply remove the node from the list of stored regions, `pushing' the pseudo-particles back up on the tree. 

One advantage of our method, compared to {\scshape treecol}, {\scshape treeray} and {\scshape trevr} introduced in Section~\ref{sec:tree_rt}, is that it is completely independent of the tree-build procedure. The implementation does not require modifying the tree-build modules which are often one of the most complicated algorithms in astrophysical simulation codes, especially with parallelization. The algorithm is easily adaptable to other types of gravity trees, whether it is built top-down or bottom-up. 

In fact, the use of threshold radii in the opening criteria, as discussed at the beginning of this section, is only necessary for top-down trees whose leaf nodes are (much) larger than the individual SPH particles. The reason originates from the fact that our adaptive algorithm relies on the nodes' labels to control the subsequent tree-walks. If the leaves of the tree are \textit{not} the particles, the individual particles would not be given labels (or any pointers) in the same convention as the nodes. Whilst it is desirable to keep the ionization front on the particle level, the mismatch of labelling/indexing system between nodes and particles makes it technically challenging to include the latter into the adaptive tree-walk algorithm. The simple solution is to separate them out from the `node-climbing' system and simply impose a radius within which particles must be extracted from the leaves. For bottom-up trees, none of these treatments would be required; using equation~(\ref{eq:opening_angle}) as the initial opening criterion would be sufficient. 

\subsubsection{Smoothing length solver} \label{sec:h_solver}

To let the nodes act in lieu of the SPH particles, we need to solve for their smoothing lengths. In analogous way to the SPH formulation, we set the nodes’ resolution lengths $h_{\mathrm{node}}$ by relating them to their local number density $n_{\mathrm{node}}$, 
\begin{equation} 
    h_\mathrm{node} = h_\mathrm{fact} \ n_{\mathrm{node}}^{-1/3},
    \label{eq:h_node}
\end{equation}
where
\begin{equation} 
    n_{\mathrm{node}} = \sum_{b_{\mathrm{node}}} W(|\textbf{r}_\mathrm{node} - \textbf{r}_{b,\mathrm{node}}|,h_{\mathrm{node}})
    \label{eq:n_node}
\end{equation}
\citep[cf. e.g.][]{price12a}. Here, $W$ is the smoothing kernel (a weight function in dimensions of inverse volume) and $\textbf{r}_\mathrm{node}$ indicates the position vector of the node. The number density is evaluated via a weighted summation over its neighbouring nodes, denoted by $b_{\mathrm{node}}$. The constant of proportionality $h_\mathrm{fact}$ in equation~(\ref{eq:h_node}) specifies the relation between smoothing length and the mean local node spacing, which also controls the number of node-neighbours in its smoothing sphere. The kernel function adopted here is the $M_4$ cubic spline \citep[][B-spline family]{schoenberg46}, truncated at $2h$. This kernel has been widely used in astrophysical SPH codes and is the current default option in {\scshape phantom}. The Exact density mapping currently implemented in {\scshape cmacionize} is also formulated based on the cubic spline kernel, though extensions to other types are possible. Following the SPH convension, we hereafter abbreviate the kernel $W(|\textbf{r}_\mathrm{node} - \textbf{r}_{b,\mathrm{node}}|,h_{\mathrm{node}})$ as $W_{ab}(h_{\mathrm{node}})$.

To solve for the smoothing lengths, we developed an approach similar to that described in \citet{pricemonaghan07} and \citet{phantom18}. This can be done by simply substituting equation~(\ref{eq:n_node}) into equation~(\ref{eq:h_node}), we formulate an equation that is just a function of $h_\mathrm{node}$:
\begin{equation} 
    f(h_\mathrm{node}) = h_\mathrm{fact} \left[ \sum_{b_\mathrm{node}} W_{ab}(h_{\mathrm{node}}) \right]^{-1/3} - h_\mathrm{node} = 0, 
    \label{eq:hnode_func}
\end{equation}
with its derivative given by
\begin{equation} 
    f'(h_\mathrm{node}) = -\frac{1}{3} h_\mathrm{fact} \left[ \sum_{b_\mathrm{node}} \frac{\partial W_{ab}(h_{\mathrm{node}})}{\partial h_\mathrm{node}} \right] \left[ \sum_{b_\mathrm{node}} W_{ab}(h_\mathrm{node}) \right]^{-4/3}  - 1 .
    \label{eq:hnode_func_deriv}
\end{equation}
The key difference between equation~(\ref{eq:h_node}) and the smoothing length expression solved in {\scshape phantom}, $h_a = h_\mathrm{fact} (m_a/\rho_a)^{1/3}$ \citep[][equation 10, see explanations therein]{phantom18}, is that the latter only applies to equal mass particles, whereas the former does not. This makes equation~(\ref{eq:hnode_func}) suitable for tree nodes whose masses differ significantly from each other. Note that the derivative term ${\partial W_{ab}(h_{\mathrm{node}})}/{\partial h_\mathrm{node}}$ in equation~(\ref{eq:hnode_func_deriv}) needs to be computed alongside $W_{ab}(h_{\mathrm{node}})$ within the same loop over the neighbouring nodes.

As nodes originate from distinct tree levels, the frequent jumps in distribution render it necessary to implement multiple back-up root-finding methods. As a first attempt, the code iteratively solves for $h_\mathrm{node}$ with the Newton-Raphson method: 
\begin{equation} 
    h_\mathrm{node,new} = h_\mathrm{node} - \frac{f(h_\mathrm{node})}{f'(h_\mathrm{node})}. 
    \label{eq:hnode_newton_raphson}
\end{equation}
Following \citet{phantom18}, the iteration is performed until $(h_\mathrm{node,new}-h_\mathrm{node})/h_\mathrm{node,0} < \epsilon_{h_\mathrm{node}}$, where $h_\mathrm{node,0}$ is the initial guess and $\epsilon_{h_\mathrm{node}}$ is the tolerance level. We set $\epsilon_{h_\mathrm{node}}$ to $10^{-2}$ as opposed to the default tolerance level of $10^{-4}$ for SPH particles because, unlike the equal mass particles, we do not (and cannot) require the node's density and smoothing lengths to be used interchangeably \citep[cf.][equation 10]{phantom18}. In fact, the smoothing length here is merely a choice. Setting a higher tolerance in exchange for faster convergence in the root-finding is hence justifiable.

The number of iterations required to solve equation~(\ref{eq:hnode_func}) largely depends on the initial guess value $h_\mathrm{node,0}$. SPH particles use the Lagrangian time-derivative of their densities to make this prediction, but since the pseudo-particles are not hydrodynamically evolved, we instead estimate this using the node sizes. If we assume that its neighbouring nodes share a similar cell length on average, then its local number density is simply equal to the inverse of the node's volume, $n_\mathrm{node} = (2 s_\mathrm{node})^{-3}$. Substituting this into equation~(\ref{eq:h_node}) gives the estimated smoothing length, 
\begin{equation} 
    h_\mathrm{node,0} = h_\mathrm{fact} (2 s_\mathrm{node}). 
    \label{eq:hnode_estval}
\end{equation}
The assumption on neighbours might not apply to regions where the pseudo-particles are `steep' on the tree, such as in areas which are very distant from the ionizing sources. Nonetheless, equation~(\ref{eq:hnode_estval}) serves as a useful approximation that allows the root-finding to converge typically within 2-3 iterations with $\epsilon_{h_\mathrm{node}} = 10^{-2}$. 

For pseudo-particles whose neighbours are distributed in a highly inhomogeneous manner, the Newton-Raphson method could fail to converge. In this case, the code proceeds to apply the bisection method. We set the initial guess range to $ \{ 10^{-2} h_\mathrm{node,0},10^{2} h_\mathrm{node,0} \} $ to account for the anomalous pseudo-particles whose smoothing lengths are far away from their estimated values. Bisection methods are more computationally expensive, but convergence is guaranteed. 

In the very rare occasion where the bisection method fails too, it likely indicates that a solution to equation~(\ref{eq:hnode_func}) does not exist. Instead of terminating the simulation, we opt to simply use equation~(\ref{eq:hnode_estval}) to be its smoothing length. This method is appropriate only if the smoothing lengths do not govern the particle densities. Should the smoothing length solver dominate the computing time (e.g. having over $10^6$ pseudo-particles; see Section~\ref{sec:runtime}), the code will also switch to using equation~(\ref{eq:hnode_estval}) for all pseudo-particles, bypassing the neighbour-find procedure. It may be difficult to maintain a consistent neighbour number for each pseudo-particle, but it can still provide good estimates for the cell densities.

Note that the SPH particles extracted near the ionizing sources (within $r_\mathrm{part}$) are \textit{not} passed into this smoothing length solver as it is more convenient to retain their original smoothing lengths. However, doing so would cause the algorithm to mistake those regions as voids. The false lack of neighbours around the leaf nodes that border $r_\mathrm{part}$'s domain could let the code erroneously converge to longer smoothing lengths. To prevent this situation, we temporarily substitute the SPH particles within $r_\mathrm{part}$ with their leaves when computing the smoothing lengths. 

\subsubsection{Neighbour-find} \label{sec:neighfind}

\begin{figure}
    \includegraphics[width=3.3in]{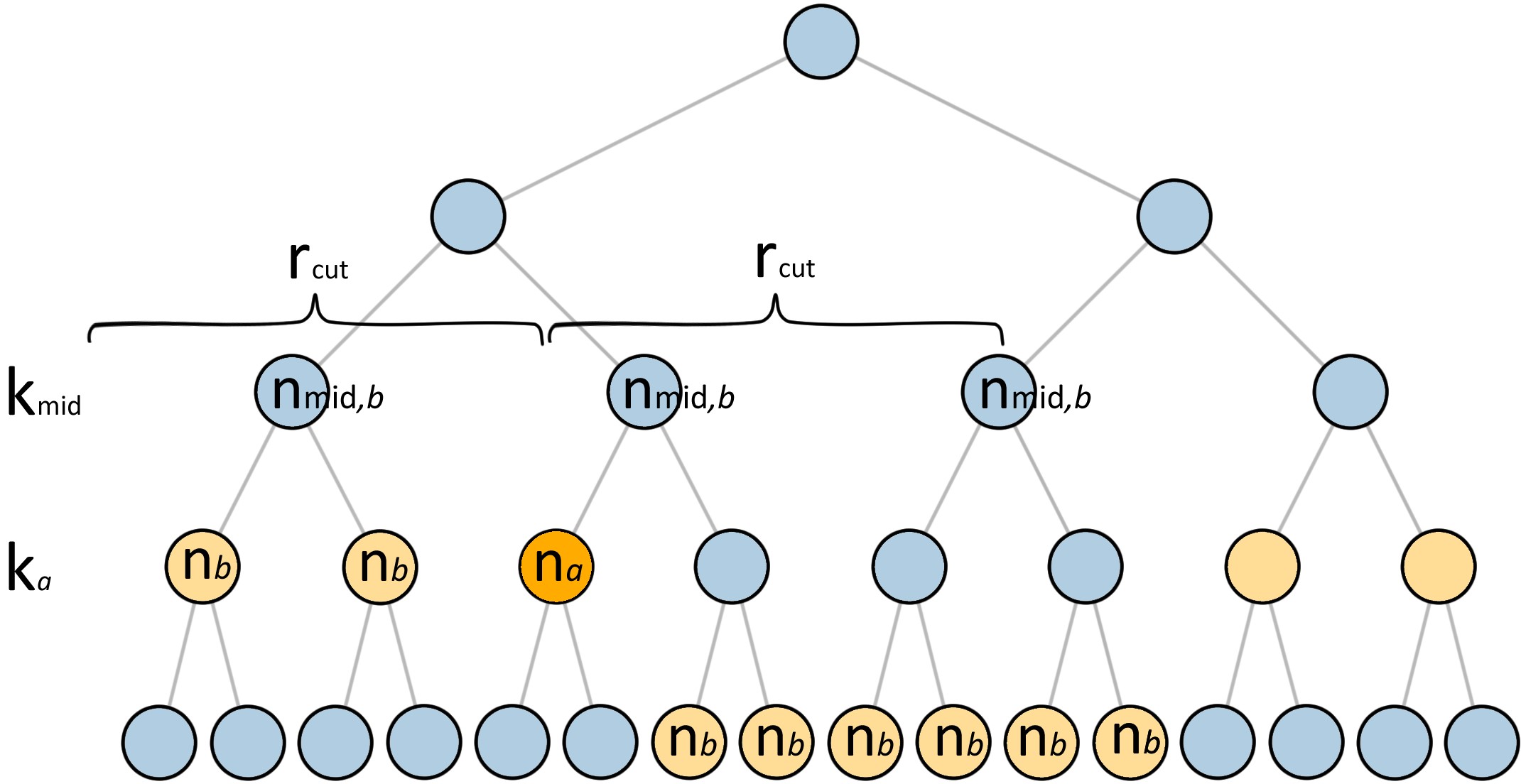}
    \caption{Illustration of the neighbour-find algorithm. Nodes in orange are example set of pseudo-particles. Consider an arbitrary target node $n_a$, coloured in darker orange, located on level $k_a$. Consider an arbitrary level above all pseudo-particles, labelled $k_\mathrm{mid}$. For nodes on $k_\mathrm{mid}$ that fall within the threshold radius $r_\mathrm{cut}$ around $n_a$, we label them $n_{\mathrm{mid},b}$; their descendant pseudo-particles $n_b$ are considered potential neighbours. }
    \label{fig:neighfind}
\end{figure}

A fast neighbour-finding scheme is vital to the smoothing length solver. For SPH particles, tree-walks are used to extract the trial neighbours, yet the fact that the pseudo-particles themselves are tree nodes precludes them from following this approach. Fortunately, the labelling system of the \textit{k}-d tree allows node relations to be recovered. We make use of this property to locate their `distant relatives'. Fig.~\ref{fig:neighfind} illustrates this algorithm. 

Consider a level somewhere midway on the tree, labelled $k_\mathrm{mid}$. Level $k_\mathrm{mid}$ must be above all pseudo-particles and, ideally, \textit{just} above the highest one. Next, consider a target pseudo-particle $n_a$. We first evaluate the distance between $n_a$ and all nodes on $k_\mathrm{mid}$, labelled $n_\mathrm{mid}$, and flag the ones that fall within a certain threshold radius $r_\mathrm{cut}$. We then loop through the rest of the pseudo-particles $n_b$. For each $n_b$, we locate its ancestor on $k_\mathrm{mid}$ using its index, which we label $n_{\mathrm{mid},b}$. If $n_{\mathrm{mid},b}$ was flagged in the previous procedure, we immediately add $n_b$ to the trial neighbour list. 

Indeed, the key for this algorithm to operate accurately and efficiently lies in setting the right threshold $r_\mathrm{cut}$. If $n_a$ is low on the tree, the size of its ancestor on $k_\mathrm{mid}$ (labelled $s_{\mathrm{mid},a}$) is usually sufficient to cover all of its neighbouring nodes. We make a more conservative estimate by taking the diagonal of the cell, giving $r_\mathrm{cut} = \sqrt{3} s_{\mathrm{mid},a}$. Of course, $n_a$ could be higher on the tree and thus closer to $k_\mathrm{mid}$, in which case the above definition of $r_\mathrm{cut}$ becomes insufficient. As such, we always perform a check afterwards by estimating the compact support radius of $n_a$, which we denote $r_{2h}$. For cubic spline kernels, this can be determined by simply taking the double of the smoothing length estimate from equation~(\ref{eq:hnode_estval}), hence
\begin{equation}
    r_{2h} = 4 \ h_\mathrm{fact} \ s_a,
    \label{eq:compact_support_radius}
\end{equation}
where $s_a$ is the size of node $n_a$. If the initially computed $r_\mathrm{cut}$ is smaller than $r_{2h}$, it likely indicates that $n_a$ is high on tree and we require an alternative method for estimating the threshold. That said, this method follows the same principle. We consider the parent of $n_a$ on one or two levels\footnote{The user may adjust this number whenever necessary, especially if the pseudo-particles are `steep' on the tree.} above $k_a$, labelling it $n_\mathrm{above}$. We again apply equation~(\ref{eq:compact_support_radius}) but replacing the node size $s_a$ by that of its parent, $s_\mathrm{above}$. This conservative approximation takes into account the jumps in tree levels amongst neighbours and we set this to be the threshold $r_\mathrm{cut}$, replacing the previous estimation.

The neighbour list is cached for fast retrieval. However, in situations where the total number of pseudo-particles is small, the overhead becomes prominent. We therefore activate this neighbour-search only if more than $10^4$ nodes (excluding the individual SPH particles but counting the leaves) are extracted from the tree; otherwise, we resolve to a brute-force approach when solving for the smoothing lengths.

\subsection{Thermal calculations} \label{sec:heating_cooling_method}

With the ionic fractions returned from the MCRT simulation, we update the internal energy of the SPH particles. For pseudo-particles that are leaf nodes or above, we make the assumption that its constituent particles, on average, share the same ionic fraction. 

RT algorithms in SPH typically directly assign fully ionized particles with a temperature of $10^4$ K. The partially ionized particles may be given a temperature that corresponds to a fraction of that for the fully ionized particles \citep[e.g.][]{kesseldeynetburkert00,dale12b}. In the original version of this RHD scheme, particles with ionic fraction over 0.5 are immediately assigned a temperature of $10^4\ \mathrm{K}$ if it is not already hotter. This heating takes place instantly, regardless of the timestep or the gas particles' properties. We hereafter refer to this as the \textit{instant} method. To ensure physical correctness, one would need to be cautious of the simulation timestep in making sure that it is consistent with the time-scales over which ionization takes place. 

This instant method is widely employed in radiative models and has been proven to be an appropriate treatment for H {\scshape ii} regions as far as dynamics is concerned. Yet, despite ionization time-scales being typically shorter than the dynamical time-scale, it is difficult to assume that thermal equilibrium must be reached by the next hydro step. Another numerical issue is that `resetting' particle energies effectively implies the removal of other thermal processes taking place in the simulation. If, for example, a supernova detonates within the H {\scshape ii} region, since cooling is necessary for modelling the supernova’s thermal input, having multiple competing methods for updating the gas internal energy operating in parallel with each other could result in a fault in the code’s thermal calculations. 

For this reason, we implement the photoionization heating and the radiative cooling, such that their contributions are added on top of the other thermal processes. This allows the internal energy of the particles to be consistently evolved free of any imposed physical assumptions, which, at the same time, improves the numerical elegance and the code architecture. 

\subsubsection{Heating} \label{sec:heating_method}

{\scshape cmacionize} computes photoionization heating rates using the accumulated path lengths of the photon packets in each cell \citep[e.g.][]{vandenbrouckewood18}. Whilst it is possible to import the heating rates from {\scshape cmacionize} to {\scshape phantom}, a simpler way is to make use of the fact that {\scshape cmacionize} is used in this scheme only as a time-independent ionization code that solves for the steady-state. At ionization equilibrium, the heating rate's dependence on the mean intensity of the radiative source $J_\nu$ cancels out, leaving with a term that represents the amount of excess energy created per photoionization event. Assuming that this energy is carried by the emitted electron, we may equate it to the kinetic energy of ideal gas. With this, for a pure hydrogen composition, the heating rate equation becomes 
\begin{equation} 
    G(H) = n_e n_p \alpha_A(H^0,T) \frac{3}{2} k T_i  \qquad [\mathrm{erg \ cm^{-3} \ s^{-1}}]
    \label{eq:heatingrate_ionequil}
\end{equation}
\citep[cf.][]{osterbrock74}, where $n_e$ and $n_p$ are the number density of free electrons and protons respectively, $k$ is the Boltzmann constant, $T_i$ is the initial temperature of the newly created electron after photoionization, and $\alpha_A(H^0,T)$ is the case-A recombination coefficient for hydrogen as a function of temperature. 

For pure hydrogen, we can write $n_e = n_p = f_\mathrm{ion} n_H$, where $f_\mathrm{ion}$ is the ionic fraction and $n_H$ is the number density of hydrogen atoms. In our code, we define number density as $n_H = \rho/m_H$, where $\rho$ is the particle density and $m_H$ is the proton mass. The electron temperature $T_i$ is computed in different ways depending on the frequency spectrum of the ionizing source. If a monochromatic source is used, we apply the equation $h(\nu - \nu_0) = \frac{3}{2} k T_i$ to obtain $T_i$, where $\nu_0$ is the minimum frequency for ionizing hydrogen ($h\nu_0 = 13.6\ \mathrm{eV}$). For sinks with a blackbody spectrum, $T_i$ can be approximated with the stellar effective temperature $T_\ast$ as long as $k T_\ast < h\nu_0$ \citep[][]{osterbrock74}. If the user request a blackbody source but only the ionizing flux is provided, we apply equation~(\ref{eq:temp_ionflux}) to estimate $T_i$. 

The case-A recombination coefficient $\alpha_A$ corresponds to the rate of collision, per unit volume, summed over recombinations (electron-captures) to all hydrogen states including the ground state\footnote{We use case-A if we assume that the recombined electrons emit secondary ionizing photons, otherwise we use case-B which excludes the captures to the ground state.}. Since $\alpha_A \propto T^{-1/2}$ \citep[see][]{osterbrock74}, by fitting the tabulated data in \citet[Table 2.1]{osterbrock74}, we formulate the expression
\begin{equation} 
    \alpha_A = 6.113 \times 10^{-11} T^{-1/2} - 1.857 \times 10^{-13} \qquad [\mathrm{cm^3 \ s^{-1}]}.
    \label{eq:caseArecomb_temp}
\end{equation}
The same method is applied to the case-B recombination coefficient, which is used if the diffuse field is not being modelled in the simulation. The expression reads
\begin{equation} 
    \alpha_B = 4.417 \times 10^{-11} T^{-1/2} - 1.739 \times 10^{-13} \qquad [\mathrm{cm^3 \ s^{-1}]}.
    \label{eq:caseBrecomb_temp}
\end{equation}

Aside from photoionization heating, we follow \citet{koyamainutsuka00} to include an extra background heating term. These heating rates, as originally calculated by \citet{wolfire95} for local Galactic conditions, cover the heating from external X-rays and cosmic rays, photoelectric heating from dust grains, the local far-UV field, and the heating by molecular hydrogen formation/destruction. \citet{koyamainutsuka02} demonstrated that, on Galactic average, these contributions can be approximated with just a constant to serve the purpose, 
\begin{equation} 
    \Gamma_0 = 2 \times 10^{-26} \qquad [\mathrm{erg\ s^{-1}}] .
    \label{eq:background_heating}
\end{equation}
This heating term $\Gamma_0$ is applied to all particles in the molecular cloud even if they are not photoionized. Note that, unlike equation~(\ref{eq:heatingrate_ionequil}), the heating rate notation $\Gamma$ quoted in equation~(\ref{eq:background_heating}) is in dimensions of energy per time. To allow for conversion, we define the relation $\Gamma = G(H)/n_H$. Hence, the total amount of heating being applied to the particles can be expressed as $\Gamma = G(H)/n_H + \Gamma_0 + \Gamma_\mathrm{shock} + \Gamma_\mathrm{PdV}$. The last two terms represent the shock heating and the heating due to gas compression respectively, which are included by default in {\scshape phantom}. 

\subsubsection{Cooling} \label{sec:cooling_method}

\begin{figure}
    \includegraphics[width=3.3in]{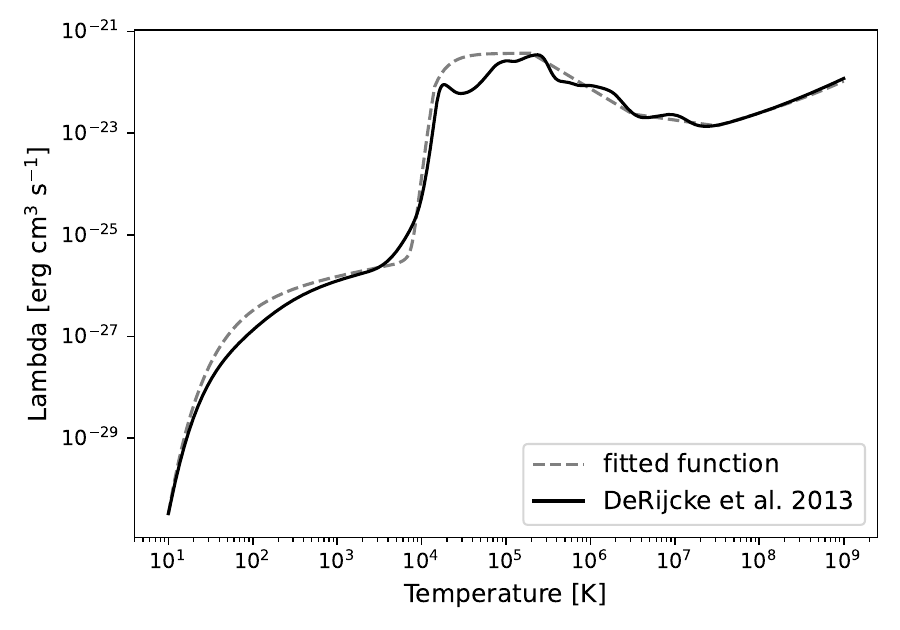}
    \caption{Cooling curve compiled by \citet{derijcke13} for density $\rho = 2.13 \times 10^{-24} \ \mathrm{g\ cm^{-3}}$, iron abundance $\mathrm{[Fe/H]} = 0.0$, magnesium-iron ratio $\mathrm{[Mg/Fe]} = 0.0$, and redshift $z = 0$. The analytic fit function (dashed line) smooths the cooling curve.  }
    \label{fig:coolcurve}
\end{figure}

The radiative cooling is implemented via a cooling curve, which is a set of pre-computed cooling rates tabulated as a function of density and temperature. This cooling rate encapsulates the effect from various atomic processes, including recombination cooling (free-bound), energy loss from collisionally excited line radiation (bound-bound), energy loss from inverse Compton scattering, and Bremsstrahlung radiation (free-free). Here, collisional ionization equilibrium is assumed. We adopt one of the cooling curves compiled by \citet{derijcke13}. In line with our current assumptions, we use the curve for iron abundance $\mathrm{[Fe/H]} = 0.0$, magnesium-iron ratio $\mathrm{[Mg/Fe]} = 0.0$, and redshift $z = 0$. The gas density assumed in this curve is $2.13 \times 10^{-24} \ \mathrm{g\ cm^{-3}}$, and the cooling rates are in units of $\mathrm{erg\ cm^{-3}\ s^{-1}}$. Since cooling rate scales with number density squared, dividing the tabulated rates by the square of its number density gives a cooling function $\Lambda(T)$, in units of $\mathrm{erg\ cm^3\ s^{-1}}$. The function is plotted in Fig.~\ref{fig:coolcurve} and it is akin to the other radiative cooling curves used in the literature \citep[e.g.][]{joungmaclow06,gent13}, as expected. 

One of the strengths of this cooling curve is that its cooling rates are consistently computed for temperatures ranging from $10\ \mathrm{K}$ to $10^9\ \mathrm{K}$. This is particularly necessary if the simulation resolves both the cold molecular gas and the shock-heated gas due to feedback, which is what this RHD scheme is designed for. Often in feedback simulations, cooling functions that are compiled by different studies are `stitched' together. For example, the cooling curve used in \citet{joungmaclow06} is comprised of the cooling rates computed by \citet{dalgarnomccray72} for $T < 2 \times 10^4\ \mathrm{K}$ and \citet{sutherlanddopita93} for $T \ge 2 \times 10^4\ \mathrm{K}$. Consistency may be hampered if different physical assumptions are used in the two papers. 

Additionally, to simplify the procedures involved in the internal energy calculations (see Section~\ref{sec:update_IE}), we fit an analytic function to the cooling curve of \citet{derijcke13}, as shown in Fig.~\ref{fig:coolcurve}. This analytic function is modified from the cooling function proposed by \citet{koyamainutsuka02}, which, similar to our purpose, was formulated to reproduce the general features of the cooling curve extensively calculated in \citet{koyamainutsuka00}. 

\subsubsection{Evolving internal energies} \label{sec:update_IE}

We update the internal energy $u$ for the subsequent hydro step via an \textit{implicit} method. Unlike \textit{explicit} methods, where the heating and cooling rates are added on to a $\dot{u}$ term in the integrator, updating the energies implicitly prevents the timestep from being overly constrained by the cooling \citep[cf.][equation 72, 288]{phantom18}. The implicit method was originally proposed by \citet{vazquezsemadeni07}, who applied it to the \citet{koyamainutsuka02} cooling function with the aim to avoid the dense shock fronts from severely restricting the timestep. Later, \citet{bonnell13} and \citet{lucas20} extended this algorithm to account for the thermal instability and the multi-phase nature of the gas pressure (and temperature), which is particularly crucial for modelling the cooling behaviour in the shock-heated regime $(T > 10^6\ \mathrm{K})$. Our algorithm builds upon these developments by adapting it to our heating terms and cooling functions. 

\begin{figure*}
    \includegraphics[width=6.0in]{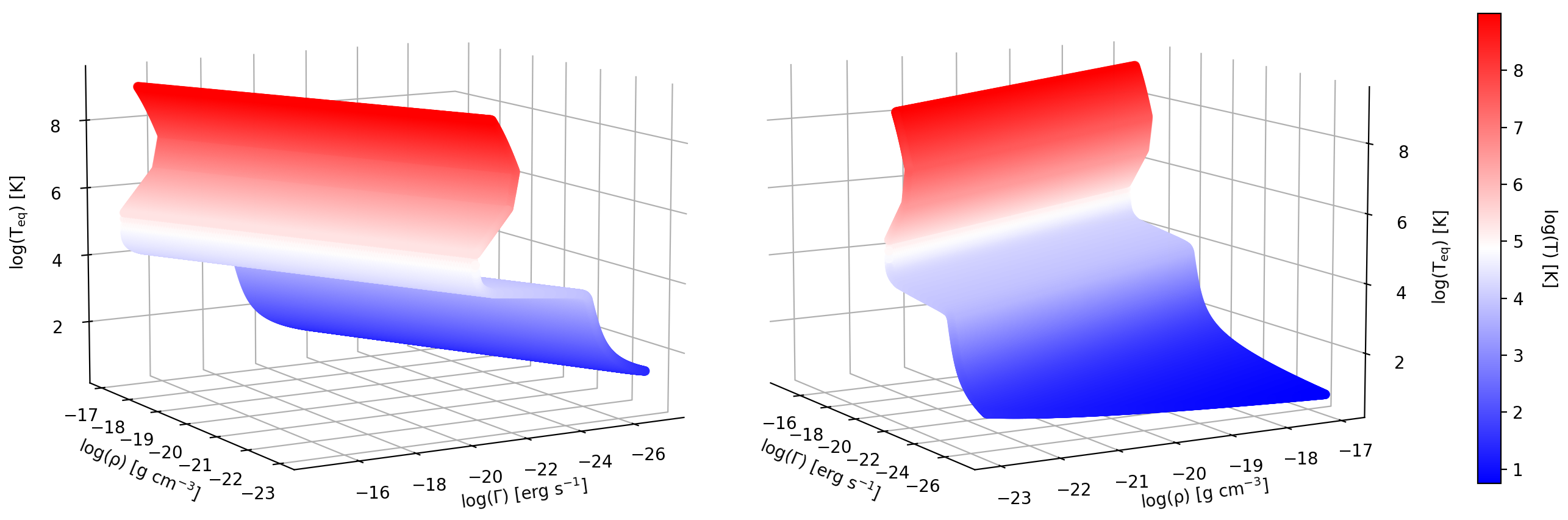}
    \caption{3-D surface plot showing the equilibrium temperatures $T_\mathrm{eq}$ as a function of density $\rho$ and heating rate $\Gamma$ under the cooling function shown in Fig.~\ref{fig:coolcurve}. Both plots are identical but at different viewing angles. The multi-valued nature of this solution surface indicates the presence of multiple thermal equilibriums at certain $\rho-\Gamma$ combinations. }
    \label{fig:Teq_solutions}
\end{figure*}

First, we pre-compute the temperatures for a range of densities $\rho$ and heating rates $\Gamma$ that satisfy the thermal equilibrium equation, $n_H\Gamma - n_H^2\Lambda(T) = 0$, where $\Lambda(T)$ is the cooling function. The solutions are presented as a 3-D surface plot in Fig.~\ref{fig:Teq_solutions}. During runtime, we interpolate from this plot to obtain the particle’s equilibrium temperature $T_\mathrm{eq}$. From Fig.~\ref{fig:Teq_solutions}, it can be seen that the $\rho-\Gamma$ parameter space is divided into 4 regimes: (a) one unique solution, that is, a vertical line placed in a direction normal to the $\rho-\Gamma$ plane would cross the solution surface once; (b) two solutions, where the line crosses twice; (c) three solutions, where the line crosses thrice; and (d) no solution. We hereafter label the $T_\mathrm{eq}$ solutions in ascending temperature order. Should multiple equilibriums exist, the $T_\mathrm{eq}$ to which the particle would approach depends on its current temperature. The key to note here is that the regime where cooling drops with increasing temperature (at around $10^5\ \mathrm{K} < T < 10^7\ \mathrm{K}$ on Fig.~\ref{fig:coolcurve}) is thermally unstable, and runaway heating would occur before it reaches the next stable equilibrium at higher temperatures. Thus, the second $T_\mathrm{eq}$ solution is likely unstable; most particles should be either at the first solution, or the third if they are substantially heated by shocks. The reason for fitting a smoothed analytic function to the cooling curve is such that the number of solutions is limited to a maximum of three, though modifying the algorithm to account for all oscillations along the cooling curve is possible.

\begin{figure}
    \includegraphics[width=3.3in]{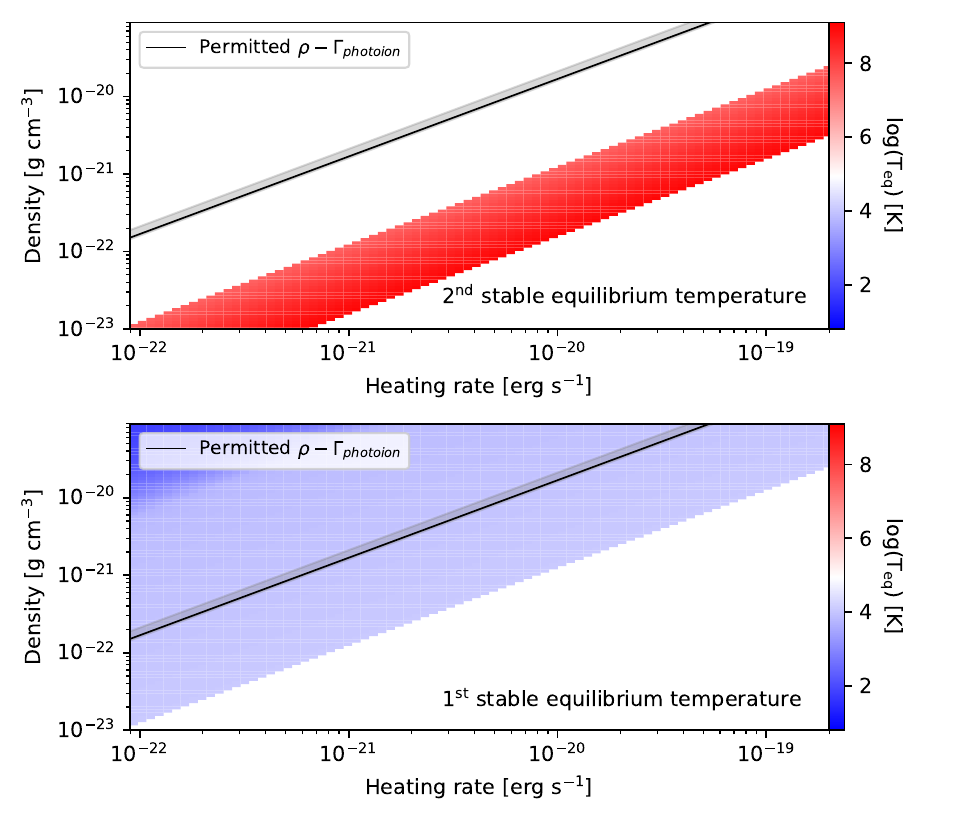}
    \caption{Top panel shows the regime in $\rho-\Gamma$ parameter space where a third $T_\mathrm{eq}$ solution (i.e. a second stable equilibrium temperature) exist. The solutions are approximately within the range $10^7\ \mathrm{K} < T_\mathrm{eq} < 10^9\ \mathrm{K}$. Bottom panel shows the regime where there is only one stable equilibrium temperature solution, which lie mostly at around $10^4\ \mathrm{K}$ or lower. Colours indicate the temperature of the solution. The black line plots equation~(\ref{eq:heatingrate_ionequil}) with $T_i = 4 \times 10^4\ \mathrm{K}$, which gives the permitted $\rho-\Gamma$ combinations under photoionization heating. The grey area marks the range for $9\times10^3\ \mathrm{K} < T_i < 6\times10^4\ \mathrm{K}$. }
    \label{fig:temp_roots_hiiregion}
\end{figure}

The temperature of H {\scshape ii} regions resulting from our heating and cooling calculations may be gauged from this surface solution. Fig.~\ref{fig:temp_roots_hiiregion} presents a `top-down view' of Fig.~\ref{fig:Teq_solutions}. The bottom panel shows, on this $\rho-\Gamma$ parameter space, the regime where there is only one $T_\mathrm{eq}$ solution ($\mathrm{1^{st}}$ stable equilibrium). These solutions lie at around $10^4\ \mathrm{K}$ or below. The top panel shows the regime where a third solution ($\mathrm{2^{nd}}$ stable equilibrium) also exists. They are seen to lie at around $10^8\ \mathrm{K}$. 

To examine whether photoionization heating can drive the gas to their $\mathrm{2^{nd}}$ stable $T_\mathrm{eq}$, we plot equation~(\ref{eq:heatingrate_ionequil}) on this $\rho-\Gamma$ parameter space, as shown by the black line, for an electron temperature $T_i$ (or stellar effective temperature $T_\ast$ if it is a sink) of $4\times10^4\ \mathrm{K}$, which is the typical effective temperature of a O7 star. This line roughly defines the permitted $\rho-\Gamma$ combinations if the gas is only subjected to photoionization. We also consider the range of $T_i$ from $9\times10^3\ \mathrm{K}$ to $6\times10^4\ \mathrm{K}$, as marked by the grey band, covering the full temperature range of OB-type stars. It is apparent that the permitted $\rho-\Gamma$ combinations from photoionization do not intersect with the $\mathrm{2^{nd}}$ stable $T_\mathrm{eq}$ regime regardless of gas density. Likewise, it is well within the boundaries of the $\mathrm{1^{st}}$ solution. This indicates that if photoionization is the only heating source present in the simulation, the ionized gas would not remain at high temperatures close to $10^8\ \mathrm{K}$, but would be correctly driven to approximately $10^4\ \mathrm{K}$.

Finally, we update the internal energy of each particle with
\begin{equation} 
    u_{i+1} = u_\mathrm{eq} + (u_{i} - u_\mathrm{eq}) \ \exp \left( - \frac{\mathrm{d}t}{\tau} \right),
    \label{eq:implicit_new_u}
\end{equation}
where $i$ denotes an arbitrary hydro step, and $u_\mathrm{eq}$ is the internal energy that corresponds to the particle's equilibrium temperature $T_\mathrm{eq}$. $\mathrm{d}t$ is the timestep, and $\tau$ is time-scale at which the gas particles radiate thermal excess or gain energy deficit. According to \citet{vazquezsemadeni07}, this is given by
\begin{equation} 
    \tau = \left| \frac{u-u_\mathrm{eq}}{n_H\Gamma - n_H^2\Lambda} \right|,
    \label{eq:heatcool_time-scale}
\end{equation}
which estimates the amount of time required for the particle to reach its thermal equilibrium. The internal energy is implicitly corrected with this method after each hydro step.

\section{Code test} \label{sec:results}

This section presents the results from testing this RHD scheme. We examine the benefits of adopting pseudo-particles as well as the accuracy of the newly implemented implicit internal energy calculations. 

\subsection{{\scshape starbench} tests} \label{sec:starbench}

We test our tree-based scheme against the well-established analytical solutions described in {\scshape starbench}, a set of standardized benchmarks compiled by \citet{bisbas15} for calibrating RHD codes. The tests are concerned with the D-type (Dense-type) expansion of H {\scshape ii} regions. As opposed to R-type (Rarefied-type), the initial phase, D-type refers to the subsequent phase where ionization equilibrium has been reached. D-type expansions are predominantly driven by the thermal pressure gradient between the ionized medium and the cool neutral medium. Since both grid-based codes and particle-based codes are involved in {\scshape starbench}, this benchmark is particularly suitable for validating our particle-mesh coupled RHD scheme. 

Consider a simple scenario where a radiative source with ionizing photon flux $Q$ is suddenly `switched on' in a uniform medium of density $\rho_0$, composed of pure hydrogen. {\scshape starbench} studies from this test case the evolution of the ionization front radius, at early-time (up to 0.14 Myr) and at late-time (up to 3 Myr). Early-time refers to the phase during which the pressure and temperature gradients across the ionization front are still large. The speed of the ionization front is subsonic relative to the ionized gas but supersonic relative to the neutral, forming a shock front ahead of the ionization front. The shocked gas is subsequently compressed into a dense shell. This early-time evolution is precisely described by analytical solutions. By equating the gas pressure in between the ionization front and the shock front to the ram pressure of the ambient neutral medium, one can arrive at the Spitzer solution, 
\begin{equation}
    R_\mathrm{Sp}(t) = R_\mathrm{St} \left( 1+ \frac{7}{4} \frac{c_{i}t}{R_\mathrm{St}} \right) ^{4/7}
    \label{eq:spitzer}
\end{equation} 
\citep[][]{spitzer78,dysonwilliams80}, where $ c_{i} $ is the sound speed in the ionized medium. Here, it is assumed that the temperature of the ionized gas $T_i$ is much higher than that of of the neutral $T_0$, such that the ratio $T_0/T_i$ becomes negligible in the equations \citep[cf. e.g.][]{bisbas15}. The Str\"{o}mgren radius $R_\mathrm{St}$ in equation~(\ref{eq:spitzer}), as first derived by \citet{stromgren39}, is given by 
\begin{equation} 
    R_\mathrm{St} = \left( \frac{3Q}{4 \pi \alpha(H^0,T) n_H^2} \right) ^{1/3}.
    \label{eq:stromgren_radius}
\end{equation}
This equation gives the ionization front radius once the H {\scshape ii} region reaches ionization-recombination equilibrium. Hence, it sets the initial radius upon entering the D-type phase, as seen in equation~(\ref{eq:spitzer}). However, the Spiter solution does not account for the inertia of the dense shell itself. This problem was later addressed in the work by \citet{hosokawainutsuka06}, who tackled the problem via an alternative approach, that is, to examine the equation of motion of the expanding dense shell. The authors arrived at what was later referred to as the Hosokawa--Inutsuka solution,
\begin{equation} 
    R_\mathrm{HI}(t) = R_\mathrm{St} \left( 1+ \frac{7}{4} \sqrt{\frac{4}{3}} \frac{c_{i}t}{R_\mathrm{St}} \right) ^{4/7}.
    \label{eq:hosokawa_inutsuka}
\end{equation}
The extra factor of $\sqrt{4/3}$ compared to equation~(\ref{eq:spitzer}) arises from the shell inertia. 

The late-time is when the pressure inside the ionized region begins to equilibrate with the external pressure. Whilst the Spitzer solution and the Hosokawa--Inutsuka solution predict an infinitely expanding H {\scshape ii} region, in reality the expansion should slowly come to a halt. The radius at which the H {\scshape ii} region stagnates has been derived by \citet{raga12b}, who modified the models of \citet{dysonwilliams80} using the pressure balance assumption. The equations of \citet{raga12b} successfully reproduce the late-time expansion curves obtained with numerical simulations, though a semi-empirically derived solution is still required to model the `overshoot' behaviour of the ionization front before stagnation \citep[][]{bisbas15}. 

Since the coupling aspect of this RHD scheme has already been validated with both the early-time and the late-time test \citep[][]{petkova21}, here we focus only on the former to judge whether or not using pseudo-particles hinders accuracy. We follow the setup specified in {\scshape starbench}. The ionizing source is monochromatic with energy $h\nu = 13.6\ \mathrm{eV}$ and flux $Q = 10^{49}\ \mathrm{s^{-1}}$. The neutral ambient medium has temperature $T_0 = 10^2\ \mathrm{K}$ and density $\rho_0 = 5.21\times10^{-21}\ \mathrm{g\ cm^{-3}}$, hence sound speed $c_0 = 0.91\times10^5\ \mathrm{cm\ s^{-1}}$. The particle mass is $10^{-3}\ \mathrm{M_\odot}$. Note that because the Spitzer solution assumes a large temperature contrast between the ionized and the neutral medium, we first temporarily disable the heating calculations and revert to the instant method. As specified in {\scshape starbench}, the temperature of the ionized region is fixed to $10^4\ \mathrm{K}$ with sound speed $c_i = 1.29\times10^6\ \mathrm{cm\ s^{-1}}$. We model the gas properties with an adiabatic equation of state and, aside from shock heating and $P\mathrm{d}V$ heating, we remove all other heating and cooling terms such that the ionized and the neutral mediums remain roughly isothermal. The adiabatic constant $\gamma$ is set to $1.00011$ and mean molecular weight is $1.0$. 

In optically thick regions, re-emitted ionizing photons from recombination can be assumed to be absorbed locally (known as the `on-the-spot' approximation). Hence, following {\scshape starbench}, a case-B recombination coefficient is used, which is fixed to $\alpha_B = 2.7\times10^{-13}\ \mathrm{cm^3\ s^{-1}}$. The photoionization cross-section used in the MCRT simulation is set to $\sigma_{\nu} = 6.3\times10^{-18}\ \mathrm{cm^2}$. The simulation does not include gravity, turbulence or any other external forces to isolate the effect of thermal pressure. The particles’ initial positions are set on a $128^3$ glass cube to minimize numerical noise. We evolve the particles on global timestepping with a maximum timestep of $10^{-4}\ \mathrm{Myr}$, which is then adaptively constrained by the limiters implemented in {\scshape phantom} as the simulation evolves.  

It is also worth noting that the temperature contrast between the ionized and the neutral gas particles may drive a shock which precedes the expanding ionization front. Shock capturing in {\scshape phantom} is done by applying artificial viscosity \citep[see e.g.][]{vonneumannrichtmyer50}. However, such excessive fictitious dissipation may erroneously lead to extra heating at regions immediately beyond the ionization front (which effectively acts as the piston described in \citet{noh87}). This is known as the wall-heating error \citep{noh87} and is present in all numerical shock-smearing methods -- in both particle-based \citep[e.g.][]{rosswog20} and grid-based codes \citep[e.g.][]{radicerezzolla12,springel10}, and in both Lagrangian and Eulerian coordinates \citep[e.g.][]{rider00,huikudriakov01}. Wall-heating is inevitable unless an artificial heat flux is added to dissipate the hot spot \citep[][]{noh87}. Hence, artificial thermal conductivity is crucial in our simulations, and we set $\alpha_u = 1$ \citep[cf.][equation 42]{phantom18}. 

\begin{figure}
    \includegraphics[width=3.2in]{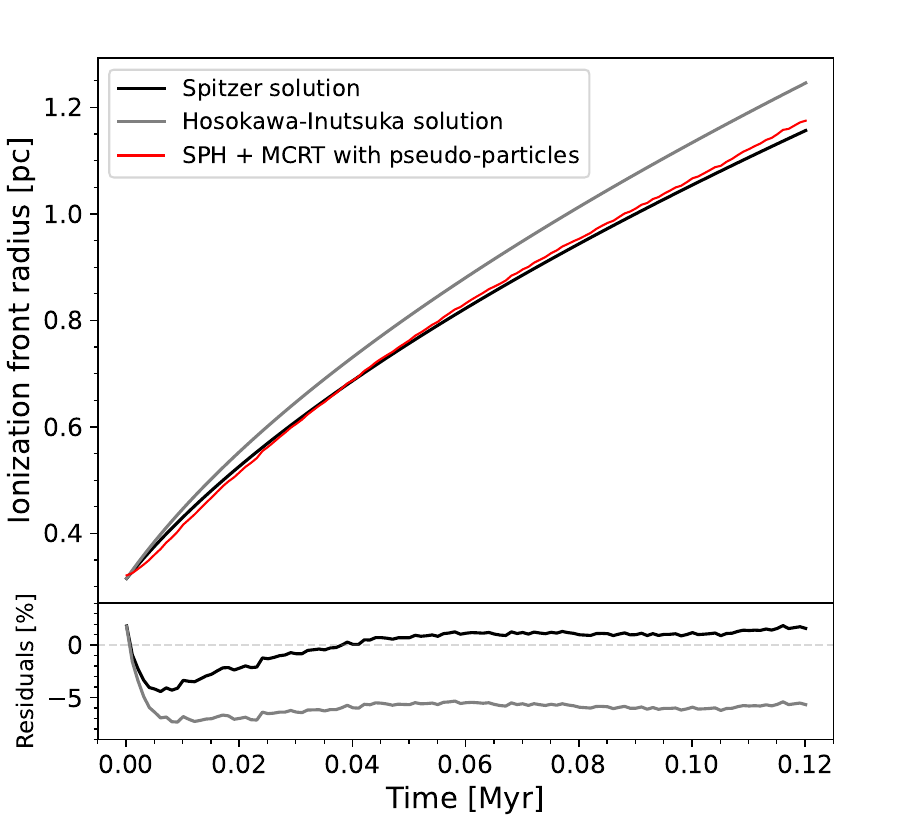}
    \caption{Top panel shows the evolution of the ionization front radius from the {\scshape starbench} early-phase test. The Spitzer solution (equation~\ref{eq:spitzer}) and the Hosokawa--Inutsuka solution (equation~\ref{eq:hosokawa_inutsuka}) are plotted in black and grey respectively. Red curve shows the results from using this RHD scheme run with pseudo-particles. Percentage errors relative to the analytical solutions, calculated with $(R_\mathrm{sim} - R_\mathrm{sol}) / R_\mathrm{sol} \times100$, are presented in the bottom panel.}
    \label{fig:ionfront_radius_evol}
\end{figure}

\begin{figure}
    \includegraphics[width=3.2in]{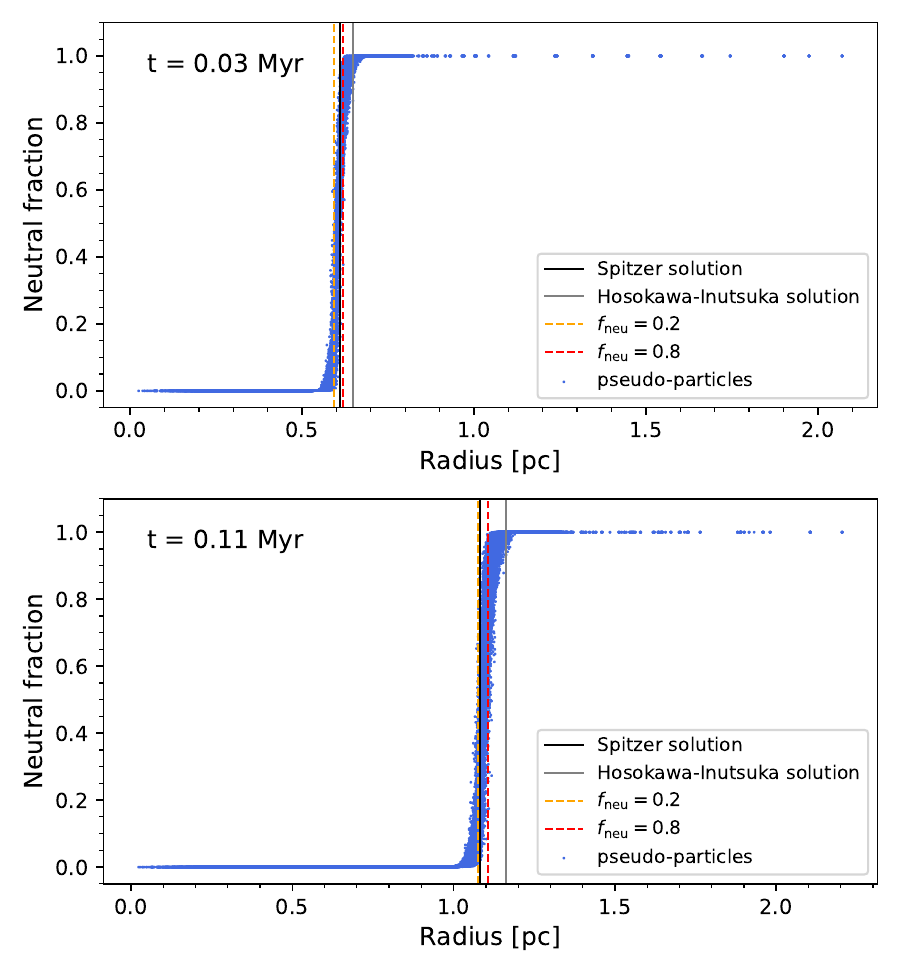}
    \caption{Neutral fraction of pseudo-particles in the {\scshape starbench} test plotted as a function of radius at $t = 0.03\ \mathrm{Myr}$ (top panel) and  $t = 0.11\ \mathrm{Myr}$ (bottom panel). The vertical dotted lines mark the radius of the shell element whose ionic fractions are closest to 0.2 and 0.8 respectively. The Spitzer solution and the Hosokawa--Inutsuka solution are shown for comparison.}
    \label{fig:ionfront_r0208}
\end{figure}

Fig.~\ref{fig:ionfront_radius_evol} shows the results comparing our tree-based RHD scheme to the Spitzer solution and the Hosokawa--Inutsuka solution. Here, regions immediately beyond the ionization front are occupied by leaves and higher-level tree nodes (see Fig.~\ref{fig:node_illus} right panels). We set $K = 100$ in equation~(\ref{eq:resol_param}) such that the ionized region is predominately resolved by individual particles. We locate the ionization front by calculating the average neutral fraction of pseudo-particles in each shell element around the ionizing source and locating the two shells whose ionic fractions are closest to 0.2 and 0.8 respectively. We show the radii of these two shells in Fig.~\ref{fig:ionfront_r0208}, plotted on top of the pseudo-particles' neutral fractions, at $0.03\ \mathrm{Myr}$ and $0.11\ \mathrm{Myr}$. The ionization front radius is defined to be the mean radius between them. 

One of the concerns regarding the use of pseudo-particles is whether or not the evolutionary models would be significantly affected by the `smearing' of ionic fractions at the partially ionized boundary, which could happen when the ionization front \textit{just} reaches the tree nodes’ domain. There may be two possible consequences: (a) the ionic fraction of a cell is lowered due to the smearing, causing it to be overlooked by the code when heating the particles\footnote{This happens with the instant method. With heating and cooling calculations, the issue may be less severe.}; or (b) once a node becomes ionized, all constituting particles heat up, causing the ionization front to erroneously accelerate. But here, we showed that our adaptive tree-walk algorithm is successful at preventing this from happening as long as the $K$ parameter (equation~\ref{eq:resol_param}) is sufficiently large (see Section~\ref{sec:tree_resolution} for a more detailed analysis). This is also thanks to the fact that our tree-walk is iteratively adjusted \textit{within} a step to eliminate the `response time', which is crucial for these time-sensitive benchmarks. Our result agrees well with the analytical solutions. It also matches the findings reported in \citet{petkova21}. The result demonstrates that using pseudo-particles immediately beyond the ionization front does not hamper the simulation accuracy. 

The reason why our expansion curve is closer to the Spitzer solution is likely because shocks in SPH are `thicker'. Whilst it is reasonable to suspect that the inertia of the shell is being underestimated in our simulations, the {\scshape starbench} test results in \citet[Fig. 8, 9]{vandenbrouckewood18} suggest an alternative explanation. The authors ran {\scshape cmacionize} in RHD mode by applying the photoionization algorithm to every hydro step. Two tests were performed - one on a static Cartesian grid, and one on a co-moving Voronoi mesh grid. It was revealed that the former gives results in strong agreement with the Spitzer solution, whereas the latter is much closer to the Hosokawa--Inutsuka solution, despite both rely on the same hydrodynamical and photoionization calculations. Comparing their density distributions, the shock in the former also appears wider. 

The findings from \citet{vandenbrouckewood18} show that H {\scshape ii} region expansion rates can be strongly influenced by the fluid model. Shocks in SPH tend to have lower peak densities and longer full-width-half-maximums due to the artificial dissipation terms as discussed earlier (see also \citet[Fig. A2]{bisbas15}). As such, the gas densities near the inner edge of the shock are higher than their counterparts in grid-codes, thus retarding the expansion of the ionization front. We note however that, as discussed in \citet{petkova21}, this issue may be addressed by increasing the SPH resolution. Alternatively, one could also recover the Hosokawa--Inutsuka solution by defining the ionization front to be the \textit{outer} edge of the transition zone (see Appendix~\ref{appen:alternative_ionfront}).

Compared to the expansion curve obtained using all SPH particles \citep[cf. e.g.][Fig. 4]{petkova21}, our curve also contains more `bumps'. This difference is more apparently seen in the residuals plot in Fig.~\ref{fig:ionfront_radius_evol}. The reason is due to the fact that tree nodes are split only along Cartesian axes during tree-build. With uniformly distributed particles, the nodes become cubically ordered. Thus, when the spherical ionized region comes in contact with the leaf nodes, the differences in node orientation relative to ionization front surface may introduce some directional inconsistencies. Though this issue has been largely addressed by the adaptive refinement, the opening of nodes may still differ slightly across the ionization front surface, leading to these minor fluctuations. 

\subsection{Runtime} \label{sec:runtime}

We now illustrate the amount of speed-up that can be achieved with our tree-based method. We run three suites of simulations with different particle numbers, $10^5$, $10^6$ and $10^7$, and vary the number of pseudo-particles at the coupling interface to measure the computing time per call to {\scshape cmacionize}. The runtime also varies with different particle/cell distributions, hence the measurement may be specific to a simulation, but the overall trend should remain similar. We begin the simulations using the uniform box from the {\scshape starbench} setup. To allow for a more realistic environment, we further impose a Gaussian turbulent velocity field with a Kolmogorov spectrum \citep[][]{kolmogorov41} on to the particles\footnote{The velocity field generation code is written by Matthew Bate (2000), based on the method described in \citet{dubinski95}.}. The turbulent gas is evolved for around $0.03\ \mathrm{Myr}$ before the ionizing source is switched on. The length of time was chosen to let the cloud develop a fractal density structure. Here, we disable the adaptive tree-walk algorithm such that the number of pseudo-particles remains constant throughout the evolution. The simulations are terminated at the $20^\mathrm{th}$-call to {\scshape cmacionize}, and we measure the mean CPU-time elapsed between each step. Error bars indicate the standard deviation across the evolutionary steps in each simulation. 

\begin{figure*}
    \includegraphics[width=6.5in]{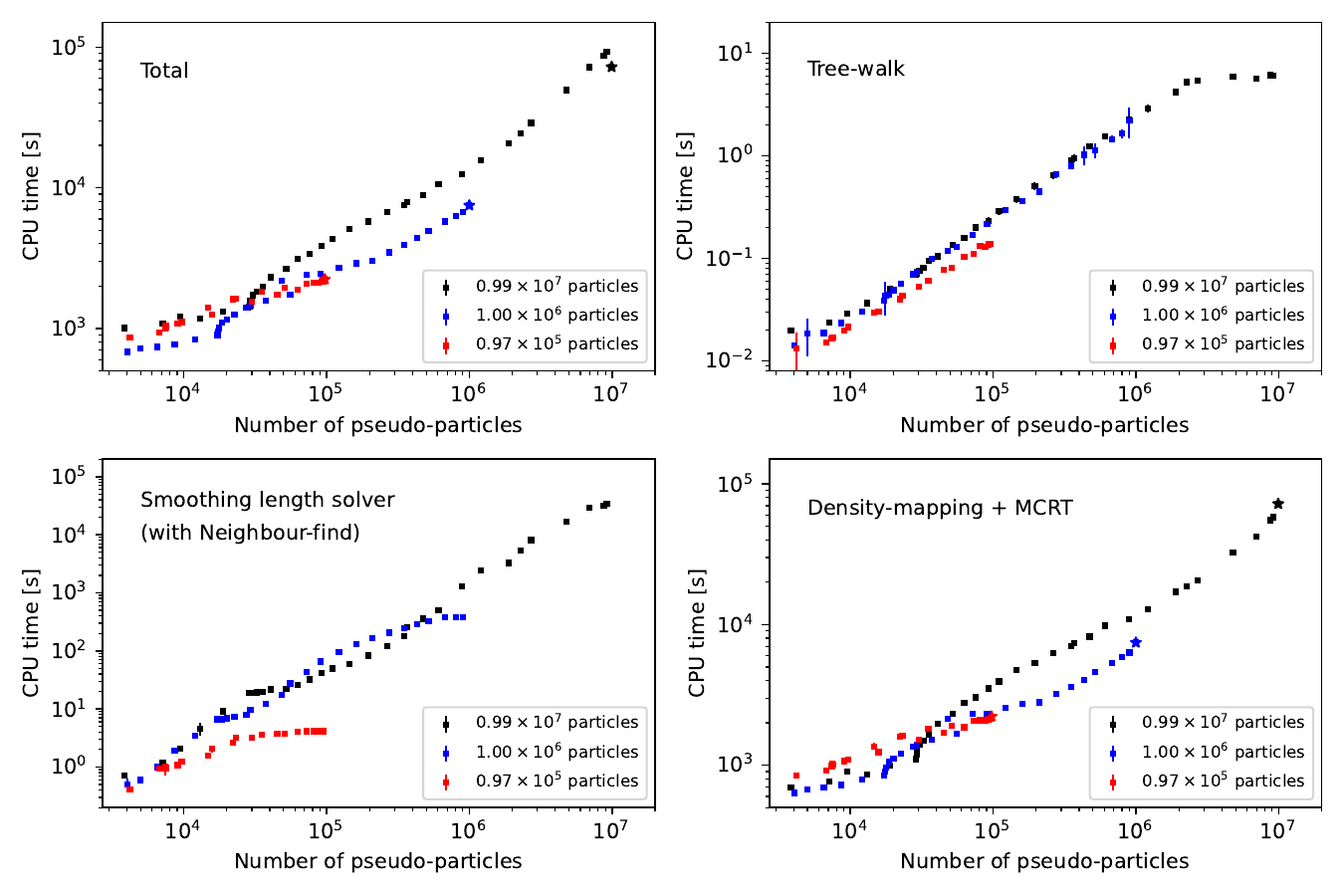}
    \caption{Variation of mean CPU time per step with number of pseudo-particles in simulations with $10^5$ (red), $10^6$ (blue) and $10^7$ (black) SPH particles in total. Top left panel shows the total runtime per step. Top right shows the computing time occupied by the tree-walk. Bottom left shows the overall time taken by the smoothing length solver (parallelized on {\scshape openmp}), including the time for neighbour-search. Bottom right shows the computing time spent on {\scshape cmacionize}'s side, covering the time for forward density-mapping, Voronoi grid construction, MCRT simulation, and inverse mapping. The star-shaped data points indicate the runs where no tree-based optimization algorithms are applied. }
    \label{fig:turbbox_runtimes}
\end{figure*}

Fig.~\ref{fig:turbbox_runtimes} shows the results, ran on sixteen $2.60\ \mathrm{GHz}$ CPU processors. We plot the total CPU time per step. We also present the runtime contributions from the tree-walk, the smoothing length solver, and the computations within {\scshape cmacionize}. The latter consists of the Voronoi grid construction, the forward density mapping, the MCRT simulation and the inverse mapping for returning ionic fractions. Since the smoothing length iteration is parallelized on {\scshape openmp}, the neighbour-find for each node is independently executed on the node's assigned thread; for simplicity, we only measure the CPU time before and after this parallel section. The $10^5$-, $10^6$-, $10^7$-particle runs are plotted in red, blue and black respectively. Each scatter point indicates the average runtime per step measured from a simulation that uses a pseudo-particle number indicated on the x-axis. We vary this by adjusting $r_{\mathrm{leaf}}$ and $r_{\mathrm{part}}$. Note that the individual particles placed around the ionizing source count towards the number of pseudo-particles. The data point at the end of each curve, marked by a star shape, indicates the run using all SPH particles, where no tree-based optimization algorithms are applied. 

The runtimes in all plots approximately follow a log-log linear relation with the number of pseudo-particles. The speed-up can be gauged by inspecting the curves. For instance, if only $10^4$ pseudo-particles are needed to keep all ionized regions resolved in a $10^7$-particle simulation, then a hundred times speed-up may be achieved. The runtime’s dependence on (pseudo-)particle distribution is reflected in the amount of deviation from a straight line. Tree-walk is the most efficient procedure, and its scaling is most unaffected by the distribution of nodes or particles. The small plateau seen at the tail of the $10^7$-particle runs likely results from our choice of $r_\mathrm{leaf}$ and $r_\mathrm{part}$, since the number of particles extracted within $r_\mathrm{part}$ varied, but the number of nodes visited by the tree-walk remained similar. 

The total computing time, in most cases, is dominated by the procedures executed within {\scshape cmacionize}, with the only exception being the $10^7$-particle runs with large pseudo-particle numbers. In this regime, the smoothing length solver becomes comparably expensive to the MCRT simulation. The neighbour-find algorithm is deemed responsible, since it involves looping over all other extracted nodes to retrieve those which lie beneath the pre-selected parents. This leads to a poor scaling with the total number of nodes extracted from the tree; future improvements on the algorithm efficiency would be desirable. In practice, however, the number of pseudo-particles would only be minimized. Hence, the issue is inconsequential, and the minimal runtime reached is our sole concern. None the less, compared to the runs using all SPH particles, the overhead of setting up pseudo-particles is close to negligible. 

\subsection{Tree resolution} \label{sec:tree_resolution}

Leaving the ionization front high up on the tree leads to error-prone simulations. But how bad is too bad? Our aim here is to gauge to what extent we can relax the checking criteria to improve the computational cost whilst maintaining a satisfactory level of accuracy. The following tests examine the effect of varying $K$ in equation~(\ref{eq:resol_param}) on H {\scshape ii} region evolution. The initial gas density is $10^{-19}\ \mathrm{g\ cm^{-3}}$, and the simulations are run with $10^6$ particles, each with mass $10^{-3}\ \mathrm{M_\odot}$. Since $K$ primarily controls the resolution of the partially ionized domains, we again apply turbulence to broaden the ionized-to-neutral transition zones along the lower density paths. To quantify non-spherically symmetrical expansions, we apply the instant heating and measure the growth in mass of gas with temperatures above $8000\ \mathrm{K}$. All simulations begin with $\mathrm{r_{leaf}} = 0.1\ \mathrm{pc}$ and $\mathrm{r_{part}} = 0.05\ \mathrm{pc}$. As the code iterates to resolve into the ionization front, if leaf nodes also fail the checking criteria, then both radii are incrementally increased by $0.01\ \mathrm{pc}$. We let the code automatically adjust the tree-walks as the simulations run. 

\begin{figure}
    \includegraphics[width=3.3in]{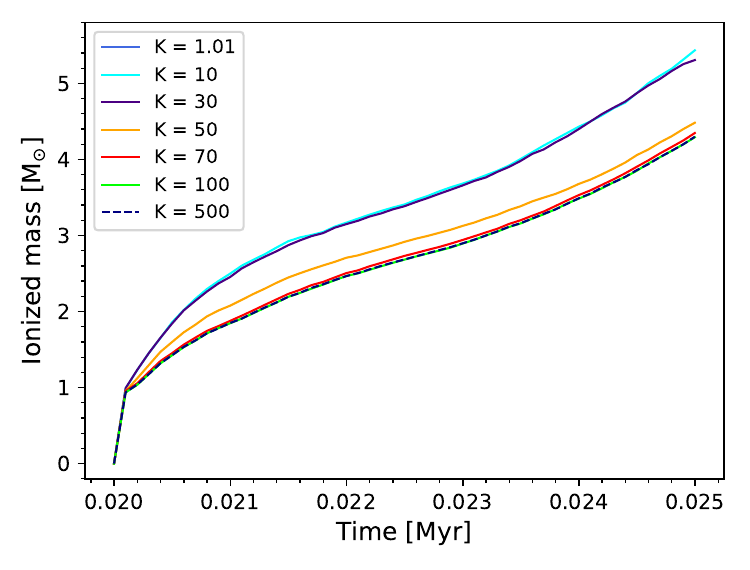}
    \caption{Growth of ionized mass in a turbulent medium with varying resolution parameter $K$ (from equation~\ref{eq:resol_param}) for the adaptive tree-walk. }
    \label{fig:ionized_mass_treeresol}
\end{figure}

The results are presented in Fig.~\ref{fig:ionized_mass_treeresol}, for $K$ values of 10, 30, 50, 70, 100 and 500. The initial ionized mass of approximately $1\ \mathrm{M_\odot}$ corresponds to the gas contained within the Str\"{o}mgren sphere. Immediately after, the curves depart from each other. The step-function-like behaviour of ionic fractions leads to this bimodal grouping of the curves, though small differences can still be seen thanks to the added density structures. The $K = 500$ curve may be interpreted as the \textit{true} values, since it almost enforces all partially-ionized domains to be on the particle level (see Fig.~\ref{fig:treeresolK_func}). For the runs with $K = 10$ and $K = 30$, plenty ionized pseudo-particles are left on the leaf level during the initial evolutionary stages, leading to approximately $1\ \mathrm{M_\odot}$ of extra masses to be erroneously heated by the end of the simulation. With $K = 50$, the results have improved, but a slight offset of around $0.2\ \mathrm{M_\odot}$ is still present. With $K = 70$ and above, the curves converge to the true value. It can be hence concluded that, for this simulation, setting $K = 70$ is sufficient to produce an accurate H {\scshape ii} region model. 

To understand by how much a coarser ionization front resolution is compensated by the computing time, we summarize the simulations' runtime in Table~\ref{tab:runtime_treeresol}. We recorded the total CPU time taken, which includes the iterating time, as well as the mean CPU time per hydro step. The results show that, in general, the runtimes scale up with the value of $K$. The anomalies seen in Table~\ref{tab:runtime_treeresol} are most likely due to the iterative calls to {\scshape cmacionize} as the pseudo-particles adjust to follow the expansion. For the $K = 10$ and $K = 30$ runs, despite being the least expensive, their accuracies are previously shown to be not within the acceptable range. Meanwhile, the $K = 70$ run is almost as accurate as $K = 500$, and furthermore, its runtime is around 19\% shorter than the latter. Compared to the run using all SPH particles, using $K = 70$ as opposed to $K = 500$ introduces an extra 9\% speed-up. It hence justifies setting a smaller value of $K$ if computing time is prioritized. 

\begin{table}
	\centering
	\caption{Runtime of simulations ran at different tree resolution parameter $K$ shown in Fig.~\ref{fig:ionized_mass_treeresol}, rounded to the nearest integer. For reference, we include the CPU time for the run with no tree-based optimization algorithm applied. }
	\label{tab:runtime_treeresol}
	\begin{tabular}{lrr} 
		\hline
		K & Mean CPU time per step [s] & Total CPU time [s] \\
		\hline
		10 & 822 & 324946 \\
		30 & 1117 & 461999 \\
		50 & 3673 & 1093908 \\
            70 & 3228 & 1148510 \\ 
            100 & 3690 & 1462631 \\
            500 & 3873 & 1417028 \\
            tree off & 8282 & 2989667 \\ 
		\hline
	\end{tabular}
\end{table}

\begin{figure*}
    \includegraphics[width=7.0in]{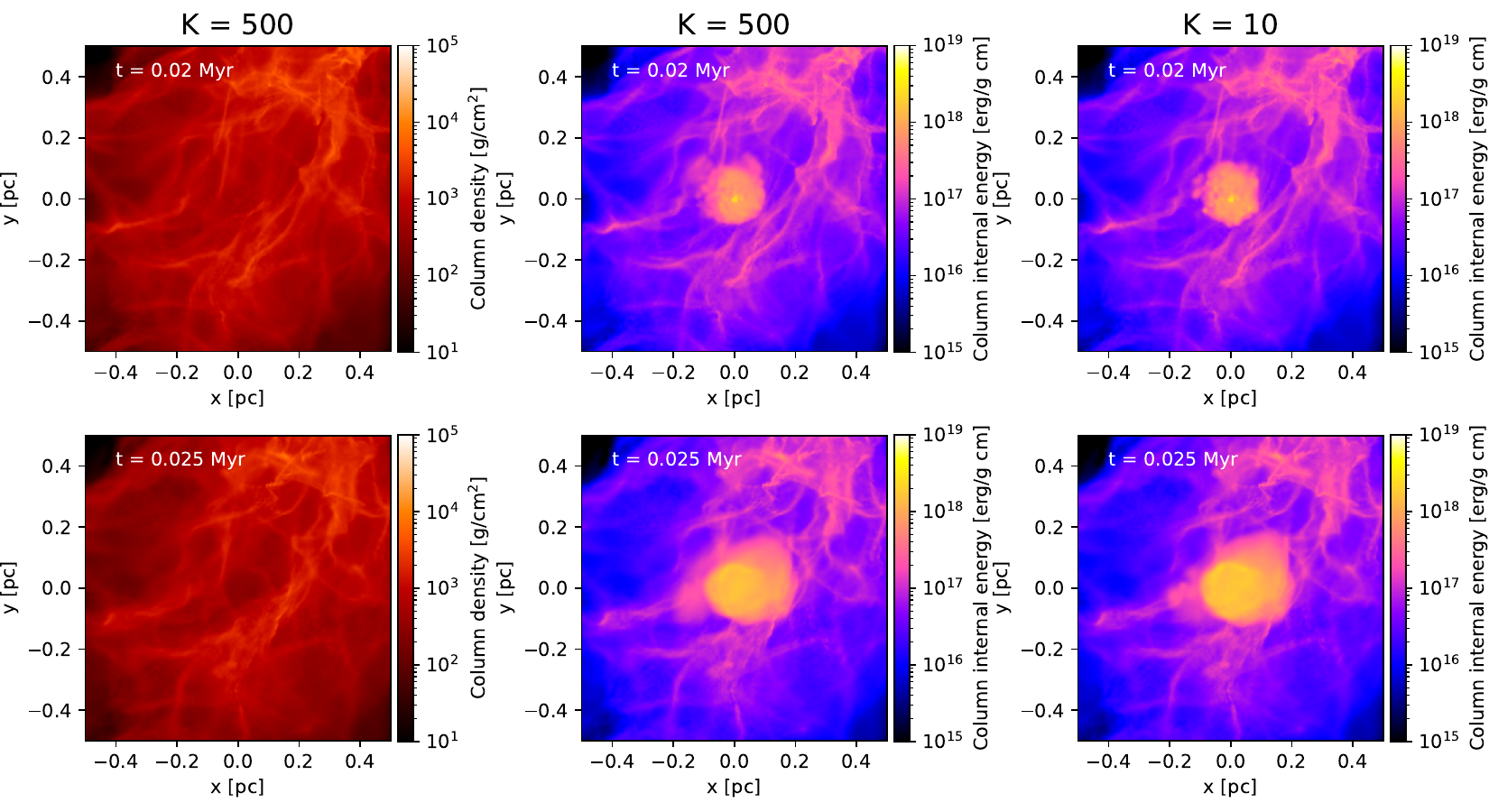}
    \caption{Column density (left panels) and density-weighted column internal energy (middle and right panels) of the turbulent medium simulation at $t = 0.02\ \mathrm{Myr}$ (top panels), immediately after switching on an ionizing source placed at the origin, and at $t = 0.025\ \mathrm{Myr}$ (bottom panels), when the H {\scshape ii} region has expanded. Left and middle panels show the run with tree-walk resolution parameter set to $K = 500$, and right panels show the run with $K = 10$, where the ionization front is left on slightly lower resolutions. }
    \label{fig:coldens_colu_treeresol}
\end{figure*}

We also inspect the morphology of the H {\scshape ii} regions resulted from the different $K$ values. In Fig.~\ref{fig:coldens_colu_treeresol}, we show the density-weighted column internal energy for $K = 10$ and $K = 500$. Additionally, we plot the gas column density of the simulation with $K = 500$ to illustrate the impact of this H {\scshape ii} region on the surrounding structures. The snapshots in the top panels are taken shortly after the ionizing source is switched on, whereas the bottom panels show the evolved H {\scshape ii} regions after $0.025\ \mathrm{Myr}$. The growth in size is apparent even within such short time interval. To the contrary, the differences between the run with $K = 500$ and that of $K = 10$ are much less prominent. During the initial times, the boundary of the heated region in $K = 10$ appears more `bumpy', since all particles within the larger ionized leaf nodes are heated, leading to a less discretized border. At later stages, the heated region in $K = 10$ is slightly larger than the $K = 500$ run, but the two are morphologically alike. It appears that loosening the criterion on ionization front resolution does not lead to entirely flawed simulation results. 

Overall, the best choice of $K$ depends on the science goals of the simulation. If the study requires a precise measurement of, for example, the effect of H {\scshape ii} region shells on the kinematics of the surrounding gas, setting $K = 500$ (or at least $K = 100$) would be necessary. However, if one only needs a rough estimate of the collective impact from multiple H {\scshape ii} regions, setting a lower $K$ may be a plausible choice. Nevertheless, to avoid creating cubes of hot gas in the simulation when $K$ value is small, the more elegant solution would be to compute the ionic fraction of each SPH particle via an interpolation from its neighbouring nodes. This would require a fast algorithm to locate pseudo-particles whose smoothing spheres overlap with the SPH particle; such developments would be ideal for future versions of this code. 

\subsection{Thermal treatments} \label{sec:therm_treatment}

In this section, we test the newly implemented heating and cooling routines. We showed in Fig.~\ref{fig:temp_roots_hiiregion} that, under the effect of photoionization, the gas particles would eventually reach their equilibrium at around $10^4\ \mathrm{K}$. What distinguishes between these particles is the time-scale over which they are heated. If the simulation timestep is shorter than the time-scale of the net heat gain, we would expect the H {\scshape ii} regions to heat up at a slower rate compared to those in the {\scshape starbench} tests. In fact, this time-scale is density-dependent. Since $\Gamma = G(H)/n_H$ and $G(H) \propto n_H^2$ (cf. equation~\ref{eq:heatingrate_ionequil}), substituting it into equation~(\ref{eq:heatcool_time-scale}) gives $\tau \propto n_H^{-2}$. This relation indicates that the lower the density, the longer it takes to reach the equilibrium temperature and vice versa. 

\begin{figure}
    \includegraphics[width=3.2in]{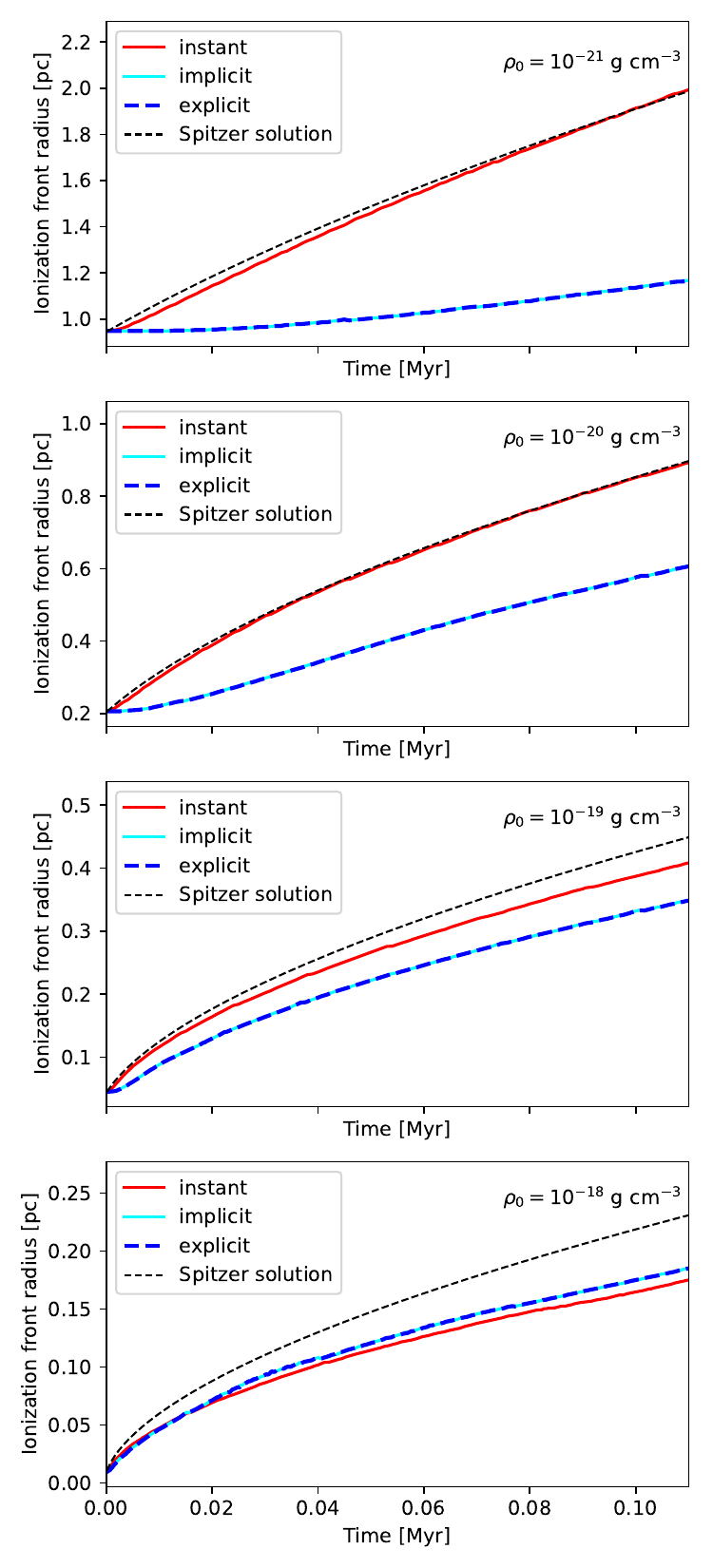}
    \caption{Expansion of ionization front radius under the instant internal energy update method (red), the explicit heating method (dark blue), and the implicit heating method (cyan). Each panel shows the result from simulation run with ambient medium density $10^{-21}\ \mathrm{g\ cm^{-3}}$, $10^{-20}\ \mathrm{g\ cm^{-3}}$, $10^{-19}\ \mathrm{g\ cm^{-3}}$, and $10^{-18}\ \mathrm{g\ cm^{-3}}$. The Spitzer solution (black) is plotted for comparison. }
    \label{fig:ionrad_therm}
\end{figure}

The effect on the expansion of ionization front is investigated. We repeat the {\scshape starbench} tests (Section~\ref{sec:starbench}) but activate the heating and cooling computations to see how much it differs from the results in Fig.~\ref{fig:ionfront_radius_evol}. The Spitzer solution assumes a large temperature gradient between the ionized and the neutral medium, and that this physical condition holds immediately after switching on the ionizing source. With heating and cooling applied, this is no longer enforced. To vary the heating time-scale, we perform the tests with initial gas densities $10^{-21}\ \mathrm{g\ cm^{-3}}$, $10^{-20}\ \mathrm{g\ cm^{-3}}$, $10^{-19}\ \mathrm{g\ cm^{-3}}$, and $10^{-18}\ \mathrm{g\ cm^{-3}}$. For each density, the simulation is run with the instant method, the explicit method, and the implicit method, as described in Section~\ref{sec:update_IE}. Because of the cooling, the temperature of the neutral ambient medium is no longer fixed at $10^2$ K; their equilibrium temperatures become a function of density (cf. Fig.~\ref{fig:Teq_solutions}), ranging from approximately $10\ \mathrm{K}$ to $10^2\ \mathrm{K}$. We set up the particles’ initial positions on a $128^3$ glass. $K = 500$ is set for the tree-walks. Fig.~\ref{fig:ionrad_therm} shows the results. 

\begin{figure}
    \includegraphics[width=3.2in]{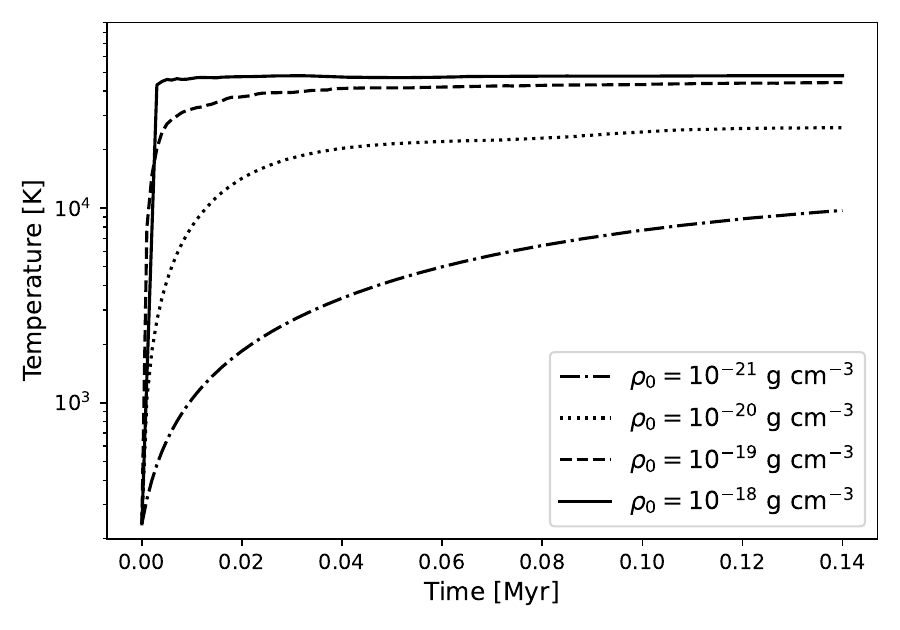}
    \caption{Mean temperature of the fully ionized particles over time under the implicit heating scheme for runs with ambient medium density $10^{-21}\ \mathrm{g\ cm^{-3}}$, $10^{-20}\ \mathrm{g\ cm^{-3}}$, $10^{-19}\ \mathrm{g\ cm^{-3}}$ and $10^{-18}\ \mathrm{g\ cm^{-3}}$. }
    \label{fig:hiitemp_rho}
\end{figure}

In all plots, the implicit curves are in perfect agreement with the explicit, hence validate the calculation method presented in Section~\ref{sec:heating_cooling_method}. Investigating the stability of this algorithm with larger timesteps would be ideal in a future study. We also see that, at lower densities, the ionization front is indeed expanding at a much slower rate compared to the instant curves as well as the Spitzer solution. The reason can be seen in Fig.~\ref{fig:hiitemp_rho}, where we plot the time-evolution of the mean temperature inside the ionized regions. At $\rho = 10^{-21}\ \mathrm{g\ cm^{-3}}$, the gas particles within the Str\"{o}mgren sphere are heating up slowly, and reach their equilibrium temperatures only towards the end of the simulation. But as density increases, at $\rho = 10^{-18}\ \mathrm{g\ cm^{-3}}$, the equilibrium temperature is reached within a couple of timesteps. This explains why, in Fig.~\ref{fig:ionrad_therm}, that the implicit/explicit curves become increasingly close to the instant curves in the high-density runs. Meanwhile, all curves begin departing from the Spitzer solution. This is likely because with a denser ambient medium, the thermal pressure within the H {\scshape ii} region is increasingly counterbalanced by the external ram pressure.  

One could also notice from Fig.~\ref{fig:ionrad_therm} that, at $\rho = 10^{-18}\ \mathrm{g\ cm^{-3}}$, the implicit/explicit curves overtake the instant curve after approximately $2\ \mathrm{Myr}$. This is in fact due to the difference in ionic fraction threshold adopted in each method. With the instant method, only the particles with ionic fractions over 0.5 are heated, whereas with heating rate computations, the term $G(H)$ is applied as long as the ionic fraction is above zero (cf. equation~\ref{eq:heatingrate_ionequil}). Hence, the latter heats more particles in the transition zone between the fully ionized and the fully neutral, causing the ionization front to expand slightly faster.

\section{Other applications} \label{sec:applications}

\subsection{Radiation pressure} \label{sec:rad_pressure}

The use of MCRT methods is not limited to ionization modelling. Radiation pressure from massive stars play an equally important role especially at the early times of the star-forming regions \citep[e.g.][]{lopez11}. The radiation pressure released in polar directions from the accretion disc may launch outflows and create cavities that influence the star formation efficiencies \citep[][]{kuiper12}. Compared to moment-based methods such as FLD \citep[cf.][]{krumholz09}, it was revealed that frequency-dependent RT methods are more suitable for modelling the pressure forces from direct stellar irradiation \citep[][]{kuiper12}, making MCRT an ideal option. 

Dynamical coupling of MCRT to hydrodynamical simulations has been developed by e.g. \citet[]{nayakshin09,noebauer12,harries19}. These RHD schemes compute the radiation force acting on each fluid element by, for example, tracking the momentum dump from the photon packets, or by applying Monte Carlo estimators to the beam energy. Alternatively, some operate by continuously transferring weighted packet momentum on to the neighbouring fluid elements. All of these methods produce smooth distributions of radiation force per unit volume, or radiation energy density, that fall radially from the radiative sources \citep[cf. e.g.][Fig. 1-3]{harries15}. 

The fact that radiation pressure impact decrements gradually over large distances, as opposed to the step-function-like behaviour of ionization, arguably makes it even more suitable for our tree-based method. It is entirely possible to swap ionic fraction in the adaptive tree-walk algorithm with radiation force per volume, or radiation energy density at the pseudo-particle, to be the node opening criterion. A smooth radially decaying pressure takes better advantage of the hierarchically built tree and likely operates well with just the opening angle (equation~\ref{eq:opening_angle}). The algorithms described in Section~\ref{sec:pseudo_particles} can be applied to this modelling problem.

\subsection{Gridless MCRT} \label{sec:gridless_mcrt}

Gridless MCRT methods that are solved directly on SPH density fields may also benefit from our tree-based optimization algorithms. In fact, most of these methods already incorporate the gravity tree for retrieving particles whose smoothing volumes are pierced by the packet trajectories (i.e. rays) \citep[][]{altay08,forgan10}. Column density is evaluated via line integrals of SPH density over the path length where the ray intersects the kernel. A notable example is {\scshape traphic}, developed by \citet{pawlikschaye08}, that computes RT on SPH with Monte Carlo approaches. This scheme employs \textit{virtual particles}, whose properties are interpolated from the neighbouring SPH particles, serving as additional sampling points in the fluid to aid the RT. In a sense, this is equivalent to superimposing a grid on to the particles, but without modifying the resolution. Despite the algorithms presented by e.g. \citet{nayakshin09,forgan10} do not rely on such techniques, we suspect that directional biases may arise if the distribution of SPH particles around the ionizing sources are highly anisotropic, such as near the feedback-driven cavities. In these situations, the transport could be preferentially directed towards the centre of mass of its neighbours \citep[see explanation in][Appendix A]{pawlikschaye08}. Thus, in general, RT algorithms that are solved locally on SPH fields can (and should) involve more computation procedures to compensate for the lack of fluid sampling points in voided regions where radiation-impacted gas mostly gathers\footnote{In our RHD scheme, this issue is addressed via the Lloyd’s iteration applied to the Voronoi grid, which improves the resolution in the voided regions.}. 

To optimize these algorithms, placing pseudo-particles at the interface between the SPH fluid model and the RT calculations is a plausible option. There are two possible methods: (a) adopt pseudo-particles in regions that are shielded or less affected by radiation; (b) for rays that lack intersecting SPH particles, go up the tree for a level or two and use pseudo-particles’ smoothing volumes to evaluate the optical depths. The resolution lengths would be increased, but this method improves the directional sampling and addresses the biases as discussed in \citet{pawlikschaye08}. In any case, the RT calculations likely need to be iterated in a similar manner to our adaptive algorithm, such that the tree-walks are adjusted according to the actual radiation field. 

\section{Summary and Future work} \label{sec:conclusion}

This paper has presented a hybrid radiation hydrodynamics scheme that couples particle-based SPH to grid-based MCRT, as first developed by \citet{petkova21}. SPH is particularly advantageous for star formation and feedback modelling, while Monte Carlo methods are known for their ability to treat radiation in highly inhomogeneous environments. In this scheme, the hydrodynamics are computed with the SPH code {\scshape phantom}. At each hydro step, the fluid densities are passed to the code {\scshape cmacionize} to run an MCRT simulation on a Voronoi grid, with generation sites (initially) coinciding with the particles’ positions. The MCRT iteratively solves for the steady-state ionization balance in each cell. The resulting ionic fractions are subsequently returned to the SPH code, with which we compute a heating term, alongside cooling, and apply it to the particle internal energies. The method we use to transfer fluid properties between the SPH particles and the Voronoi cells is the Exact density mapping method of \citet{petkova18}. However, whilst these numerical methods are all highly accurate, the computational cost can be immense. 

Our solution is to temporarily reduce the fluid resolutions at the interface between the SPH and the MCRT code using the tree-based gravity solvers. Gravity trees are often combined with RT computations to achieve efficient RHD schemes. We developed an adaptive tree-walk algorithm to transform tree nodes into pseudo-particles in regions that are less affected by the ionizing radiation. To let the pseudo-particles act in lieu of SPH particles, we developed a smoothing length solver and a neighbour-find algorithm, both of which are dedicated to tree nodes. The tree-walks are iteratively adjusted according to the ionic fractions returned from the MCRT code, allowing the pseudo-particles to adapt to the evolving radiation field. This algorithm is independent of the tree-build procedure, making it suitable for all types of gravity trees. The use of pseudo-particles is not limited to ionization modelling, and, in fact, can be applied to couple SPH to many other numerical methods. 

Extending the code to treat atomic species other than hydrogen would be desirable. In that case, the heating contributions from helium and other metals would need to be computed with Monte Carlo estimators in the MCRT code. Transferring temperatures and heating rates between the two codes becomes necessary. Cooling curves for higher metalicities may be imported from \citet{derijcke13}. Alternatively, full chemistry calculations may be performed to model the cooling. Another beneficial extension to this scheme is to make full use of MCRT simulations, that is, to account for dust grains in the model. As far as photoionization is concerned, absorption by dust reduces the Str\"{o}mgren radius and thus lowers the expansion rates \citep[e.g.][]{haworth15}. But if radiation pressure is incorporated, momentum deposition may be amplified \citep[e.g.][]{agertz13}, and the expansion of H {\scshape ii} regions could speed up if the shell is optically thick to IR \citep[][]{krumholzmatzner09}. Either way, the effect of dust on H {\scshape ii} region evolution warrants further investigation.

\section*{Acknowledgements}

We thank the anonymous reviewer for the very insightful questions and suggestions that helped improved the paper; and {\scshape phantom} \citep[][]{phantom18} and {\scshape splash} \citep[][]{splash07}. CSCL thank all the helpful advice on code development from James Wurster, and his comments on the SPHERIC2024 conference proceeding, the precursor of this paper. CSCL also thank Thomas Bisbas, Ruben Cabezón and Ahmad Ali for the discussions, as well as Bert Vandenbroucke for the technical help with {\scshape cmacionize}. This work was supported by the STFC training grant ST/W507817/1 (Project reference 2599314). The simulations were performed using the HPC Kennedy, operated by University of St Andrews. The SPH figures in this paper were created using the python package {\scshape sarracen} \citep[][]{sarracen23}.

\section*{Data Availability}

The data associated with this paper, including the simulations codes and outputs, can be made available upon reasonable request to the corresponding author. 



\bibliographystyle{mnras}
\bibliography{ref_list} 




\appendix

\section{Tree resolution parameter} \label{appen:tree_resol_K}

\begin{figure}
    \includegraphics[width=3.3in]{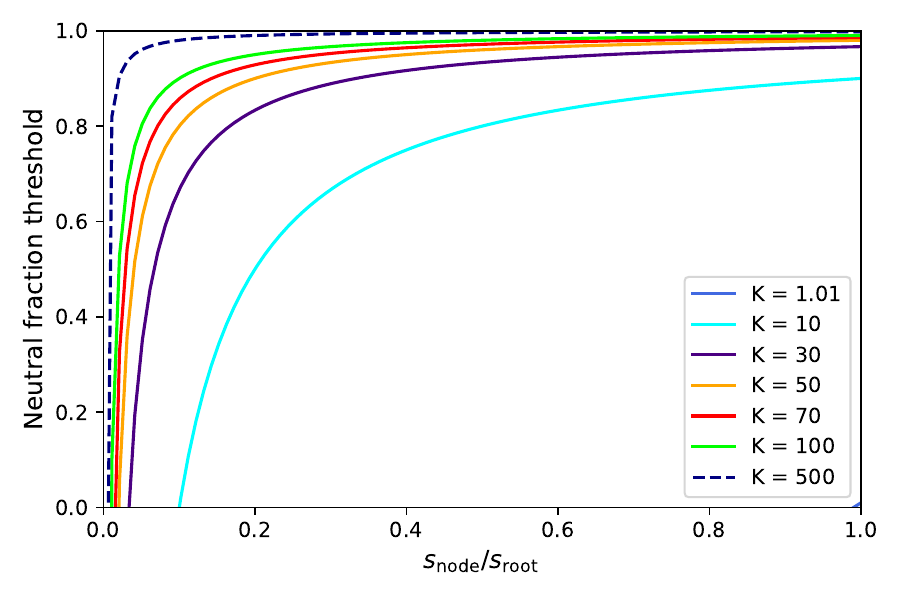}
    \caption{Plot of equation~(\ref{eq:resol_param}), the node opening criterion for the adaptive tree-walk, for $K = 1.01$, $K = 10$, $K = 30$, $K = 50$, $K = 70$, $K = 100$ and $K = 500$. The x-axis denotes the size of the node relative to that of the root. The curves define their corresponding lower limit in neutral fraction. }
    \label{fig:treeresolK_func}
\end{figure}

The opening criterion in the adaptive tree-walk system aims to ensure that all ionized nodes are sufficiently small in size. We parametrize this criterion via equation~(\ref{eq:resol_param}). Fig.~\ref{fig:treeresolK_func} plots this function for $K = 1.01$, $K = 10$, $K = 30$, $K = 50$, $K = 70$, $K = 100$ and $K = 500$. The threshold in neutral fraction determines whether or not a node needs to be opened in the next tree-walk during the iterative calls to the MCRT simulation. Take $K = 30$ as an example. If a pseudo-particle of size $s_\mathrm{node}/s_\mathrm{root} = 0.4$ has an ionic fraction of 0.2 (i.e. a neutral fraction of 0.8), the code considers it under-resolved, and the node will be opened in the subsequent tree-walk. This reduces the node size by approximately half and shifts the pseudo-particle(s) towards the left on Fig.~\ref{fig:treeresolK_func}. This process is repeated until all pseudo-particles lie above the $K = 30$ curve, such that those with high ionic fractions are sufficiently close towards the left on this plot. Thus, the parameter $K$ controls the strictness of the opening criterion. Setting $K = 1.01$ (seen at the lower right corner) effectively switches off the adaptive algorithm. On the other hand, with $K = 500$, the function approaches both axes of the plot. This means that as long as the neutral fraction drops slightly below 1.0, all nodes must be opened, and the ionized region must be resolved by the individual SPH particles. 

\section{Alternative ionization front} \label{appen:alternative_ionfront}

\begin{figure}
    \includegraphics[width=3.3in]{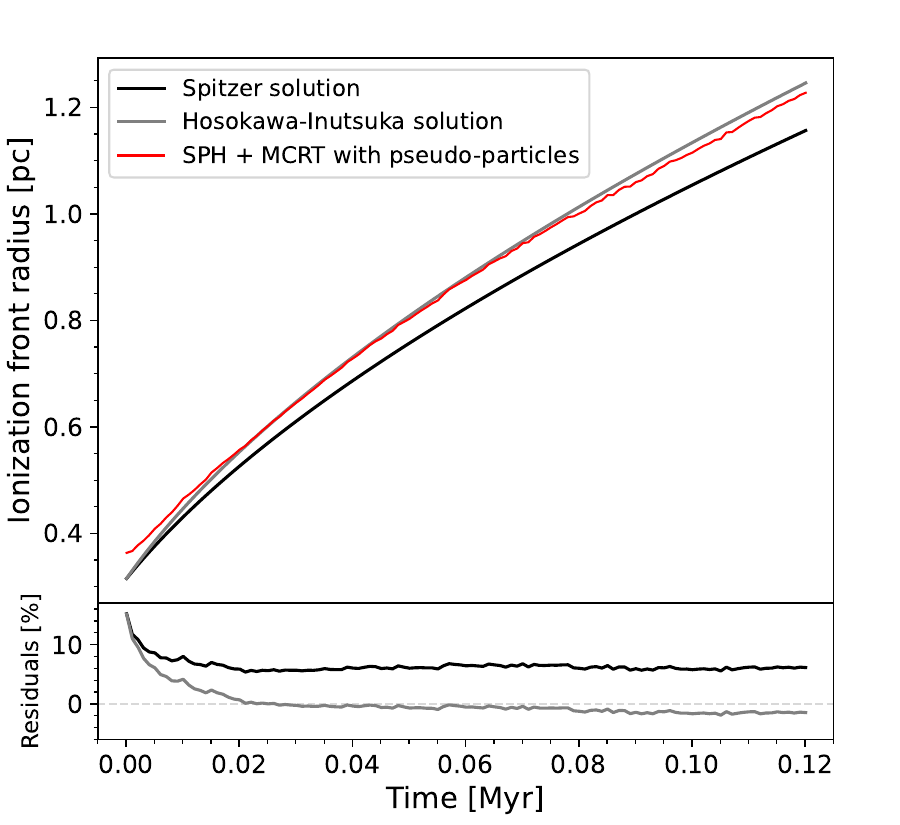}
    \caption{Same as Fig.~\ref{fig:ionfront_radius_evol} but defining the ionization front to be the midpoint between the radii that correspond to ionic fractions of 0.01 and 0.02 respectively, instead of 0.2 and 0.8. }
    \label{fig:ionfront_radius_evol_r098099}
\end{figure}

The original {\scshape starbench} test \citep[][]{bisbas15} defines the ionization front to be the radius at which the ionic fraction is equal to 0.5. Likewise, in this paper and in \citet{petkova21}, the ionization front is defined to be the midpoint between the two radii that correspond to ionic fractions of 0.2 and 0.8. In fact, the analytic theories \citep[][]{spitzer78,hosokawainutsuka06} assume a thin shell approximation and do not consider the width of the shock, nor distinguish between the shocked shell and the ionization front. These assumptions introduce some ambiguity in comparing the solutions to simulations. Under the current definition, Fig.~\ref{fig:ionfront_radius_evol} shows that our expansion curve is closer to the Spitzer solution. However, from Fig.~\ref{fig:ionfront_r0208}, the Hosokawa--Inutsuka solution appears to be tracing the outer edge of the transition zone. If we instead define the ionization front to be the midpoint between the radii of ionic fractions 0.01 and 0.02, we obtain the expansion curve shown in Fig.~\ref{fig:ionfront_radius_evol_r098099}. Due to the uncertainty in the definition of ionization fronts in the analytical models, it may be argued that our RHD scheme also reproduces the Hosokawa--Inutsuka solution. 

\bsp	
\label{lastpage}
\end{document}